\documentclass[bibyear]{aa}
\usepackage[usenames,dvipsnames]{xcolor}

\usepackage{graphicx}
\usepackage{amsmath, bm}

\usepackage{tikz}
\usepackage{txfonts}
\usepackage{longtable}
\usepackage{listings}
\usepackage{color}
\usepackage{textcomp}
\usepackage{multirow}
\usepackage[normalem]{ulem}
\usepackage{soul}
\usepackage{amsmath}

\usepackage{natbib,twoopt}
\usepackage[breaklinks=true]{hyperref} 
\bibpunct{(}{)}{;}{a}{}{,}             
\makeatletter
  \newcommandtwoopt{\citeads}[3][][]{\href{http://adsabs.harvard.edu/abs/#3}%
    {\def\hyper@linkstart##1##2{}%
     \let\hyper@linkend\@empty\citealp[#1][#2]{#3}}}
  \newcommandtwoopt{\citepads}[3][][]{\href{http://adsabs.harvard.edu/abs/#3}%
    {\def\hyper@linkstart##1##2{}%
     \let\hyper@linkend\@empty\citep[#1][#2]{#3}}}
  \newcommandtwoopt{\citetads}[3][][]{\href{http://adsabs.harvard.edu/abs/#3}%
    {\def\hyper@linkstart##1##2{}%
     \let\hyper@linkend\@empty\citet[#1][#2]{#3}}}
  \newcommandtwoopt{\citeyearads}[3][][]%
    {\href{http://adsabs.harvard.edu/abs/#3}
    {\def\hyper@linkstart##1##2{}%
     \let\hyper@linkend\@empty\citeyear[#1][#2]{#3}}}
\makeatother

\newcommand{\msun}{\mbox{$\,\mathrm{M_{\odot}}$}}

\DeclareMathOperator{\atantwo}{atan2}

\begin{document}

\title{Mapping the anisotropic Galactic stellar halo with Blue Horizontal Branch stars}

\author{João A. S. Amarante\inst{1}, 
        Sergey E. Koposov\inst{2,3,4}, 
        Chervin F. P. Laporte\inst{1}}
    
\institute{Institut de Ciències del Cosmos (ICCUB), Universitat de Barcelona (UB), Martí i Franquès, 1, 08028 Barcelona, Spain \email{joaoant@gmail.com}
    \and{Institute for Astronomy, University of Edinburgh, Royal Observatory, Blackford Hill, Edinburgh EH9 3HJ, UK}
    \and{Institute of Astronomy, University of Cambridge, Madingley Road, Cambridge CB3 0HA, UK}
    \and{Kavli Institute for Cosmology, University of Cambridge, Madingley Road, Cambridge CB3 0HA, UK}
    }

\date{Received \today; }

  \abstract{We use Legacy Survey photometric data to probe the stellar halo in multiple directions of the sky using a probabilistic methodology to identify Blue Horizontal Branch (BHB) stars. The measured average radial density profile follows a double power law in the range $ 5 < r_{gc}/{\rm kpc} < 120$, with a density break at $r_{gc}\approx20$ kpc. This description, however, falls short, depending on the chosen line-of-sight, with some regions showing no signature of a break in the profile and a wide range of density slopes, e.g. outer slope $-5.5 \lesssim \alpha_{out} \lesssim -4$, pointing towards a highly anisotropic stellar halo. This explains in part the wide range of density profiles reported in the literature owing to different tracers and sky coverage. Using our detailed 3-D stellar halo density map, we quantify the shape of the Pisces overdensity associated with the transient wake response of the Galaxy's (dark) halo to the Large Magellanic Cloud (LMC). Measured in the LMC’s coordinate system, Pisces stands above the background, is 60 degrees long and 25 degrees wide and aligned with the LMC’s orbit. This would correspond to a wake width of $\sim 32$ kpc at $\sim 70$ kpc. We do not find a statistically significant signature of the collective response in density as previously reported in the literature measured with K giant stars, despite our larger numbers. We release the catalogue constructed in this study with 95,446 possible BHB stars and their BHB probability.
  }
\titlerunning{Legacy Survey BHB stars}
\authorrunning{Amarante, Koposov, Laporte}
\maketitle

\section{Introduction}
\label{sec:intro}
The Milky Way's (MW) stellar halo provides valuable information about the Galaxy's history since it contains most of the metal-poor, ${\rm [Fe/H]}\,<\,-1$, and oldest stars, age $>10$ Gyr \citep[e.g.][]{beers+2005,frebel+2015,das+2020}. Although it is the most extended baryonic component of the Galaxy, reaching out to $\approx$ 200 kpc, its stellar content only contributes around 2\% to the total stellar mass of the Galaxy (e.g., \citealt{deason2019}). Moreover, the stellar halo is not smooth, but is comprised of substructures from globular clusters \citep[e.g.][]{myeong+2018-gcs,massari+2019,forbes2020,limberg+2022}, to debris from past accretion events \citep[e.g.][]{helmi+99, koppelman+2019,naidu+2020,yuan+2020, limberg+2021, malhan+2022}, and satellite dwarf galaxies \citep[e.g.][]{mcconnachie2012}. Throughout the manuscript, we will refer to the Galactic stellar halo as the ``halo'' unless otherwise explicitly stated when referring to the dark matter halo of the MW.\par
Determining the structure of the MW halo involves counting stars across the sky. With the current photometric surveys, this task is limited by dust extinction, the survey's magnitude limit and its footprint, which are encoded into the survey selection function. Knowing the survey selection function is crucial for a rigorous approach to measuring the radial density profile of the halo in order to interpret observations with theoretical models. For instance, a non-smooth halo would imply a recent merging activity whereas a smooth one would reveal a more quiescent formation history \citep{johnston+2008, deason+2013}. We expect both, with varying degrees of contributions, in the hierarchical formation scenario \citep[e.g.][]{searle, White-Frenk91, springel+06}.\par
Despite the difficulties in obtaining a sample of stars that homogeneously covers the Galaxy, numerous studies mapped the halo density profile albeit with a not-well-established consensus on the shape of the profile. The radial density profile has typically been described by a single-, double-power law, or Einasto profile \citep[e.g.][]{deason+2011,xue+2015,hernitschek+2018, wu+2022} with the choice adopted partially due to the sample and survey limitations. A halo described by a double power law also defines the radius at which the inner and outer slopes change, known as the break radius, $r_{br}$. \par
\citet{deason+2013} showed that a broken radial density profile is a signature of apocenter pile-ups of accreted dwarf galaxies. They studied the radial density profiles of idealized simulations with accreted satellites from \citet{bullockjohnston2005} and argued that the presence of a break radius results from a massive merger event, i.e. a merger that contributes the most to the total stellar halo mass. They concluded that the presence of a break radius in the MW stellar halo density profile is, therefore, a signature of a massive merger event (see e.g. \citealt{gilmore2002,brook,meza2005} for other signatures of the same merger). Recently, the accreted dwarf that dominates the MW stellar halo, and likely causes the observed broken profile, was identified as the Gaia-Sausage/Enceladus \citep[GSE;][]{belo2018, helmi2018, haywood} and is estimated to constitute up to 3/4 of the inner ($r_{gc} < 30$ kpc) stellar halo (e.g. \citealt{wu+2022_gsecontrib}, but see also \citealt{lane+2023}). Interestingly, \citet{han+2022} and \citet{yang+2022}, using different methodologies, found evidence for a doubly broken halo density profile, and associated the two break radii with the {apocenters} of the GSE orbit \citep{naidu+2021_nbody}. \par
Since then, the GSE debris properties have been extensively studied taking advantage of the Gaia data combined with large spectroscopic surveys. Its stellar mass, peak [Fe/H] and time of merger range from $0.15-1\times 10^9$ \msun\, \citep[e.g.][]{feuillet+2020,naidu+2020, limberg+2022, lane+2023}, $-1.45 < {\rm [Fe/H]} < -1.17$ \citep[e.g.][]{amarantea, bird+2021, liu+2022}, 8-11 Gyr \citep[e.g.][]{gallart+2019,montalban+2021,xiang_rix2022}, respectively. The differences arise from distinct selection criteria and samples used (see e.g. \citealt{carrillo+2024} for a discussion). Despite the GSE merger event being estimated to have happened 10 Gyr ago, part of its debris is argued as being not completely phase-mixed. For instance, the Hercules-Aquila Cloud South and North (HAC-S, HAC-N, \citealt{belokurov+2007,watkins+2009,simion+2014}) and Virgo Overdensity (VOD, \citealt{vivas+2001,newberg+2002, juric+2008}) -- overdensities observed in the stellar halo -- are chemodynamically consistent with GSE stars \citep[][]{simion+2019, perottoni+2022}. These substructures, except for HAC-N, have been argued to be a consequence of the GSE merger as seen in pure N-body models \citep{naidu+2021_nbody}. On the other hand, \citet{donlon+2020, donlon+2023} argue that the presence of these substructures indicates a more recent time for the merger, circa $\sim$2--3 Gyr ago.    \par
While the GSE debris is likely to dominate the stellar halo within the break radius, at larger distances, a more heterogeneous stellar halo, built from the debris of several other minor accretions, is found. The outer halo also comprises a myriad of known dwarf-satellite galaxies bound or in the process of being engulfed by the MW \citep[e.g.][]{belokurov+2007dwarfs, koposov+2015, pace+2022, cerny+2023}. For instance, ongoing accretion events are observed for the Saggitarius dwarf (Sgr, e.g. \citealt{ibata+1994, johnston1995}), and Small and Large Magellanic Clouds (SMC, LMC, e.g. \citealt{besla+2007}). Due to their relatively massive nature, pure N-body simulations of these two systems put into question the ``Galaxy in equilibrium" approximation. The stellar content and the dark matter halo of the MW are strongly perturbed due to the interaction with Sgr \citep[e.g][]{purcell+2011, gomez+2013, laporte+2018b,laporte+2019, carr+2022, borbolato+2024}, and the LMC \citep[e.g.][]{weinberg+2006, laporte+2018a, petersen_wakesim, garavito-camargo+2019,vasiliev2023}. \par
The LMC should induce a gravitational wake in the MW dark matter distribution and cause the sloshing of the stellar halo {in a classical \citet{chandrasekhar1943-dynfric} sense from stars, and dark matter, being deflected behind the LMC's orbit.} Thus, the LMC's orbit wake should create an overdensity in the stellar halo, known as the ``transient wake" \citep{weinberg1998,laporte+2018a,garavito-camargo+2019, petersen_wakesim, erkal+2021, garavito-camargo+2021}. Besides that, the movement of the MW-LMC system barycenter \citep{gomez+2015} can also create a signature in the dark matter halo and consequently imprinted on the stellar component, namely the ``collective response" \citep{weinberg1989,weinberg1998,garavito-camargo+2019,cunningham+2020, petersen_penarrubia2021, rozier+2022}. Numerical work also predicts an overdensity of LMC-stripped stars at the track of the LMC orbit \citep{petersen_lmcstripped}.\par
Recently the Pisces overdensity \citep{sesar+2007, watkins+2009,kollmeier+2009, nie+2015} has been interpreted as a signature of the local wake on the stellar halo \citep{belokurov+2019}. This has been corroborated further with K-giants by \citet{conroy+2021} which also suggested a detection of the collective response in star counts. To further complicate matters, retrograde debris has also been associated with GSE debris in the region close to Pisces \citep{chandra+2023_echoes}.\par
Given these various contributions to the stellar halo, i.e. different accretion events and dynamical processes, a single-density profile would fall short of describing its complex structure. Ideally, by probing the whole sky with a homogeneous sample, one can measure any spatial dependence on the density profile. In fact, \citet{hernitschek+2018} has done it using RR-Lyrae from the Pan-STARRS1 survey \citep{sesar+2017} and fitted the stellar halo density profile for several bins in Galactic longitude, $l$ and Galactic latitude, $b$, with $30^\circ$ and $60^\circ$ width, respectively. After removing known overdensities, they concluded that the stellar halo is well fit either by a single power law or an Einasto profile in all directions, which at face value would be considered at odds with a broken profile and with an out-of-equilibrium MW stellar halo. \par
Probing the stellar halo at large distances depends on measuring accurate stellar distances. Currently, this is mainly done with ``standard candles". These are stars with relatively well-defined distance-module relations, such as red giant branch (RGB), RR Lyrae, and blue horizontal branch (BHB) stars. In this work, we will rely on BHB stars as they are in general associated with the MW stellar halo due to their old age \citep{dotter+2010} and low metallicity \citep{santucci+2015}. \par
BHB stars have a relatively well-defined absolute magnitude range \citep[e.g.][]{sirko+2004} and they can be identified photometrically in the colour range, $-0.3 < g-r < 0$ \citep[e.g.][]{yanny+2000, deason+2011}. Although spectroscopy is necessary to confirm BHB stars \citep[e.g.][]{vickers+2012, barbosa+2022}, an appropriate colour selection can identify BHB stars with high confidence \citep[e.g.][]{belokurov_koposov+2016,li+2019,starkenburg+2019}. \par
In this work, we use Legacy Survey data release 9 to probe the stellar halo density profile with BHB candidates. However, our measurement does not rely on identifying individual BHB stars. Similarly to \citet{deason+2011}, we developed a methodology which takes into account the survey footprint, and for a given spatial selection, we measure the BHB fraction from the colour-colour distribution in a probabilistic fashion. This allows us to robustly estimate the halo density profile with its associated errors. \par
This paper is organised as follows: in Section \ref{sec:data} we present the Legacy Survey data selection and cleaning. In Section \ref{sec:methodology} we describe the methodology to calculate the BHB star fraction at a given spatial and magnitude selection, and how we model the radial density profile. We proceed by showing the density profiles for the MW, known overdensities regions, and several patches in the sky with high angular resolution, compared to previous studies, in Section \ref{sec:results}. We discuss the implications of our findings for the structure of the MW stellar inner and outer halo in Section \ref{sec:discuss}. Finally, in Section \ref{sec:conclusion} we summarize our work and discuss future prospects. 

\section{Data}\label{sec:data}

In this section, we briefly describe the Legacy Survey, the colour selection to discriminate BHB star candidates and possible sources of contamination.

\subsection{Legacy Survey data selection}
We use data from the Legacy Survey data release 9 \citep[LSDR9][]{desi+2019}. LSDR9 {covers} a total area of 14,000 deg$^2$, in the region $-18^{\circ}<\delta < +84^{\circ}$ and galactic latitude $|b|>18^{\circ}$. The data contain measurements of the $g$, $r$, $z$ photometric bands reaching a magnitude depth of approximately 24, 23.5 and 22.5 respectively. Its photometry coverage makes it an excellent database for photometrically identifying BHB star candidates. \par
We initially select objects within the extinction corrected colour\footnote{LSDR9 uses \citet{schelegel+1998} Galactic extinction maps and provides the coefficients at \url{https://www.legacysurvey.org/dr9/catalogs/\#galactic-extinction-coefficients}.} intervals $-0.4 < (g-r) < 0.2 $ and $-0.7 < r-z < 0.1$, and within the apparent magnitude range $12 < g < 24$\footnote{All photometric magnitudes are extinction corrected thus we opted not to write them with the subscript ``$0$''.}. The first step of the data processing consisted of applying the LSDR9 bitmasks\footnote{\url{https://www.legacysurvey.org/dr9/bitmasks/}} 12 and 13 which remove pixels touching a large galaxy and globular cluster, respectively. This procedure removes objects within nearby galaxies from the Siena Galaxy Atlas 2020 \citep{sga} and stars close to globular clusters from LSDR9 catalogue. We also applied the $g,r,z$ bands offset correction given by \citet{zhou+2023}. They showed the presence of an offset between LSDR9 with Gaia synthetic photometry which is also strongly correlated with sky position. We stress the importance of this correction, as without it, any colour-based study can be severely biased.

\subsection{Colour-colour space}\label{sec:ccBHB}
\begin{figure}
\centering
\includegraphics{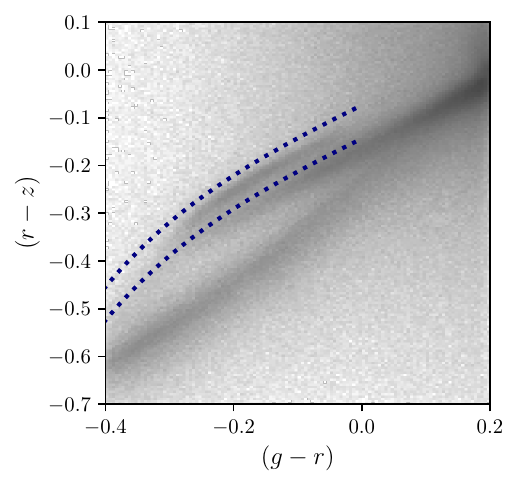}
    \caption{The $(g-r)\times(r-z)$ diagram for the first selected LSDR9 objects. The blue dotted lines are the boundaries for selecting BHB star candidates following \citet{li+2019}.}
    \label{fig:color-ridge}
\end{figure}

Figure \ref{fig:color-ridge} shows the $(g-r) \times (r-z)$ plane of our initial sample. BHB stars are well known to lie within $-0.3<(g-r)<0$ \citep[e.g.][]{deason+2011, belokurov_koposov+2016} and this is our starting point to define the BHB fraction in a probabilistic fashion. \citet{li+2019} defined the boundaries for BHB stars using $(g-r)$ and $(r-z)$ colours, and here we re-write their relation defining the $grz$ colour:

\begin{equation}
    \begin{split}
        grz = & 1.07163(g-r)^5 - 1.42272(g-r)^4 + \\
        &0.69476(g-r)^3 - 0.12911(g-r)^2 +\\
         &0.66993(g-r) - 0.11368 - (r-z),
    \end{split}
    \label{eq:bhb}
\end{equation}
where BHB stars are scattered around $grz \approx 0$.\par
As seen in
Figure \ref{fig:color-ridge}, the region associated with BHB stars, within the dashed blue lines, is within the limits of previous studies. This is reassuring as, although we use the same photometric band, the calibration of LSDR9 could be significantly different to offset the BHB relation. Compared to previous studies, we only slightly modified the range on $(g-r)$ colour. Thus, throughout the paper, we only consider stars satisfying the following colour criteria:
\begin{equation}\label{eq:cc}
\begin{split}
    &-0.14 < grz < 0.07 \\
    &-0.30 <(g-r)<-0.05
\end{split}
\end{equation}

\begin{figure*}
\centering
\includegraphics{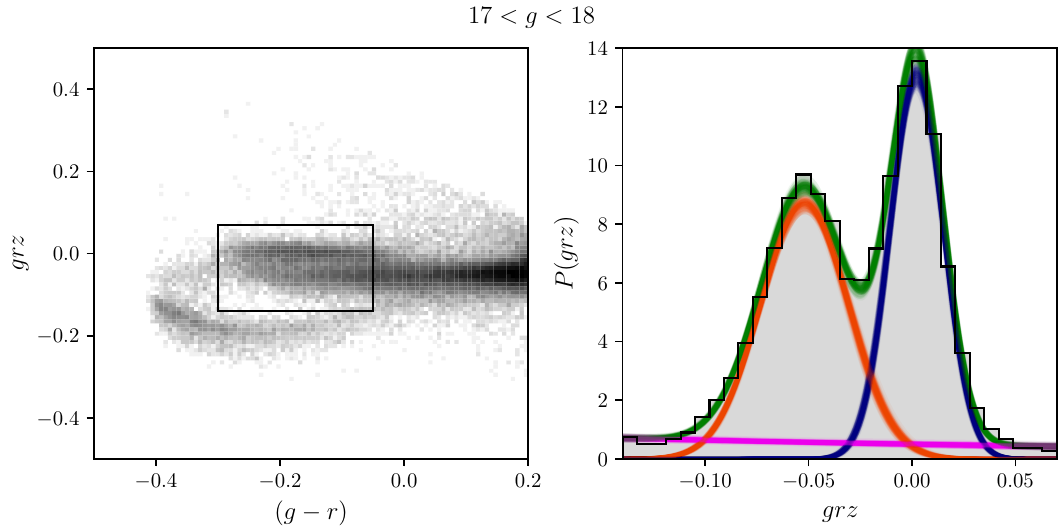}
\includegraphics{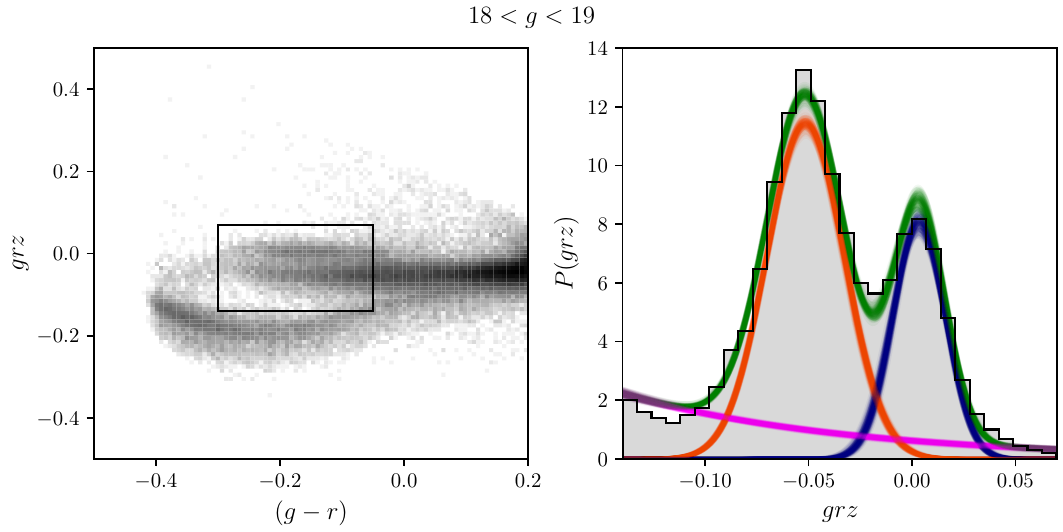}
    \caption{The left panels show $grz$ colour, see Equation \ref{eq:bhb}, as a function $(g-r)$ at different $g$ ranges. The right panels show the $P(grz)$, in grey, of the stars within the selection box on the left panel. The corresponding fit to the data following Equation \ref{eq:Pgrz} as well as the BHB, BS and background component are represented in green, blue, orange and magenta, respectively. These two examples are part of the exercise illustrated in the left column of Figure \ref{fig:mcmc-g} which assumes flat priors for the free parameters. Throughout this paper, we do use Gaussian priors for the BHB stellar component when calculating the BHB stellar densities, see Section \ref{sec:meth} for details.}
    \label{fig:color-ridge2}
\end{figure*}

\subsection{Possible sources of contamination}\label{sec:contamination}

The colour-colour selection defined in Section \ref{sec:ccBHB} can be contaminated by blue stragglers (BS), quasars, and white dwarfs \citep[see e.g][]{vickers+2012}. In the next section, we will explicitly take into account the BS contamination, and here we apply some criteria to minimize the number of quasars and white dwarfs contaminants.\par
At faint magnitudes and towards bluer colours, there is an increasing number of quasars among all the observed objects. Moreover, quasars have a large scatter over the colour space due to their different spectral energy distributions and redshifts \citep[see e.g.][]{ross+2012}. To remove potential quasar objects, we use the fluxes on the LSDR9 $r-$band and the Wide-field Infrared Survey Explorer \citep[WISE,][]{wise2010paper} photometric band $W1$, as the quasar's dust emission makes them extremely bright in the $r-$band. We include the criterion $-0.09<{\rm F}_{W1}/{\rm F}_r<0.45$ to select objects of interest, where $F_{W1}$ and $F_{r}$ are the observed fluxes in $W1$ and $r$, respectively. This selection is motivated by the fact that typical quasar objects have $F_{W1}/F_r \gtrsim 0.45$. \par
To estimate the amount of contamination from white dwarfs, we used the Gaia EDR3 white dwarf catalogue \citep{fusillo+2021} and verified that they have negligible contamination in the BHB region of this work. For instance, white dwarfs start to populate the selection box for $g>17$ but mostly in the redder part of the $grz$ colour. See Appendix \ref{app:wd} for details.\par
We also removed stars that lie within 5 times the effective radius of known dwarf galaxies from  \citet{mcconnachie2012} catalogue, except for the Large and Small Magellanic Clouds (LMC, SMC) which we removed stars within an angular separation of 12.5 degrees and 6.5 degrees from their centres, respectively. Finally, we also mask the region with $|\tilde{B}_{sgr}|<10.5^{\circ}$, where $\tilde{B}_{sgr}$ is the Sgr stream coordinate system as defined in \citet{belokurov+2014}.

\begin{figure*}
\centering
\includegraphics{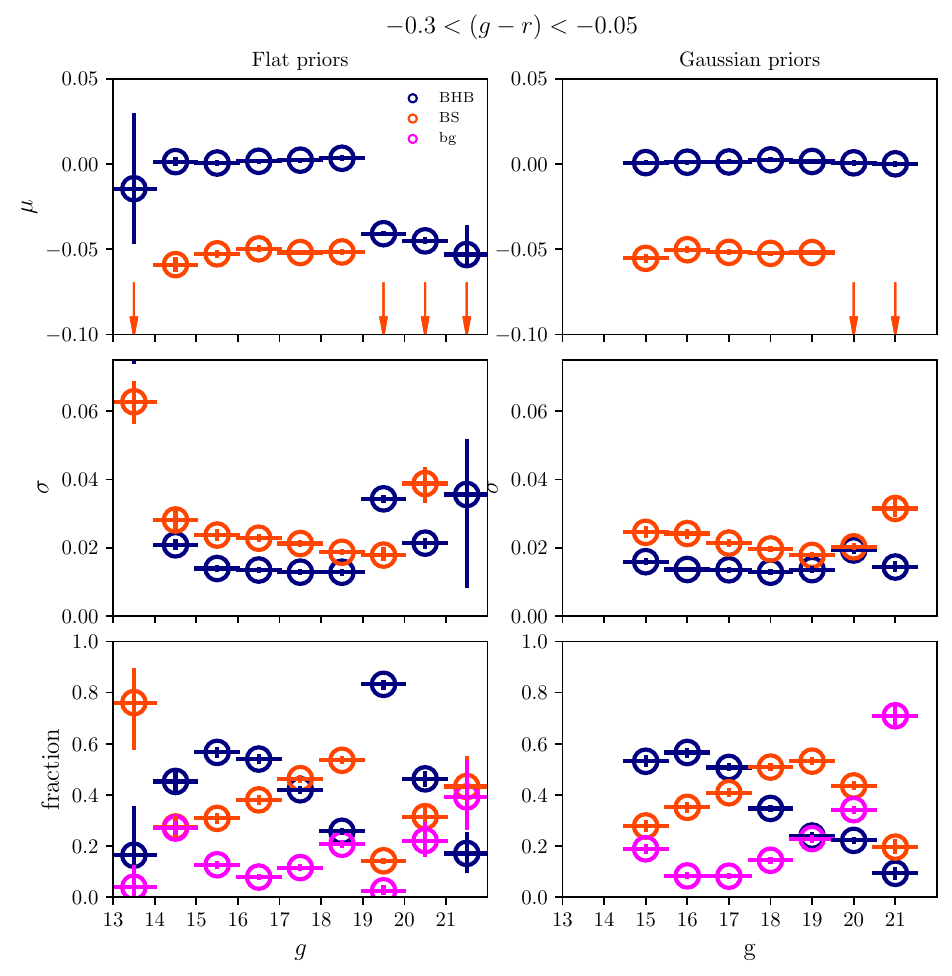}
    \caption{\textbf{Left column:} the variation of the free parameters while fitting Equation \ref{eq:Pgrz} as a function of $g$ assuming flat priors. While the horizontal lines correspond to the $g$ interval on which the fit was performed, the vertical lines correspond to the 16th-84th interval of the posterior distribution. The BHB stellar parameters are relatively constant within $15<g<19$ which motivated the use of Gaussian priors throughout this work, see Section \ref{sec:meth} for discussion. \textbf{Right column:} now with Gaussian priors on the BHB component and within $14.5 < r < 21.5$. The decrease in the BHB fraction as a function of magnitude is expected due to the intrinsic density profile of the MW halo.}
    \label{fig:mcmc-g}
\end{figure*}
\section{Methodology}\label{sec:methodology}
In the previous section, we defined the colour-colour selection, Equation \ref{eq:cc}, of the main sample which contains most of the BHB stars, and removed contamination from other sources, namely quasars and white dwarfs. We now proceed with the methodology to obtain the fraction of BHB stars.
\subsection{BHB stars fraction}\label{sec:meth}
For any given selection within our sample, the BHB stars' contribution to the $grz$ distribution can be modelled as a Gaussian component with mean $\mu=0$  and dispersion, $\sigma$, due to the errors and scatter of the relation given by Equation \ref{eq:bhb}. We illustrate this in Figure \ref{fig:color-ridge2}
where we show the $grz$ histogram for stars within two different $g$ intervals. In the first example, in the magnitude range $17 < g < 18$, the $grz$ colour distribution has a double peak shape, each associated with the BHB and BS components. Thus, we model the $grz$ probability density distribution, $P(grz)$, as a two-component Gaussian mixture with an additional exponential component representing other contaminant sources missed in the data cleaning of the previous section, i.e.:

\begin{align}
    P(grz) = w_{1} P(grz|BHB) + w_2 P(grz|BS) + w_3 P(grz|OTH)& \nonumber\\
=w_{1} G(grz,\mu_{1}, \sigma_{1}) + w_2 G(grz,\mu_{2}, \sigma_2) + w_3 \exp(\beta grz)&
\label{eq:Pgrz}
\end{align}
where $G(x,\mu_i,\sigma_i)$ is the probability density function (PDF) of the Normal distribution for the BHB and BS components with mean $\mu_i$ and standard deviation $\sigma_i$. The background exponential is characterised by the parameter $\beta$. The contribution fraction of each component is given by $w_i$, and their sum equals 1. Therefore, there are a total of eight free parameters. \par
Selecting BHB star candidates using a probabilistic model based on the star's photometry \citep[e.g.][]{deason+2014,thomas+2018, starkenburg+2019,yu+2024} or based on spectroscopic parameters \citep[][]{sirko+2004,xue+2011,santucci+2015,barbosa+2022} and model their spatial distribution has been extensively explored. However, these approaches fail to select BHB stars at the faint end of the magnitude range, where the photometric or spectroscopic uncertainties are large and the overlap between BS and BHB is significant. The methodology presented in this work lies in estimating the number (or fraction) of BHB stars at any given selection within our sample, instead of selecting individual objects based on their probabilities \citep[see, e.g.][]{deason+2011,fukushima+2019}.  \par
We estimate the number of BHB stars by fitting the distribution given by Equation \ref{eq:Pgrz} and sample the model with the Affine Invariant Markov Chain Monte Carlo (MCMC) sampler, implemented in the {\tt emcee} package \citep{emcee}. This allows us to obtain an error estimation from the posterior distribution of the eight free parameters of Equation \ref{eq:Pgrz}. For each $w_{bhb,i}$ posterior, we calculate the number of BHB stars, $N_{bhb,i}$, as:
\begin{equation}\label{eq:Nbhb}
    N_{bhb,i} = w_{bhb,i}*N_{P,i},
\end{equation}
where $N_P$ is a random number drawn from a Poisson distribution with observed objects $\lambda = N_{tot}$, $N_{tot}$ is the total number of objects within a given magnitude and/or spatial selection, and $i$ is the posterior index. From the posterior distribution, we use the 50th percentile as the value of $N_{bhb}$, and the difference between the 84th and 16th percentiles as its error\footnote{For a Gaussian distribution this relation corresponds to $1\sigma$}.  We now proceed with a few tests to verify any dependence on colours and $g$-band for the BHB Gaussian component, $(\mu_{bhb}, \sigma_{bhb})$, which could affect the modelling in the next section.  \par
\subsubsection{BHB star fraction test}
We select stars with $-0.30 < (g-r) < -0.05$ and fit the $grz$ distribution using Equation \ref{eq:Pgrz}, including the errors of $grz$, for several intervals in $g$-band.
Here we use flat priors for all the parameters. The right column of Figure \ref{fig:color-ridge2} shows two examples of the fitting. While for the intervals shown the BHB stellar component is centred around $grz \approx 0$, this is not the case for the fainter intervals, $g>19$, where the flat priors fail to provide a good fit for the data.
In fact, for the interval $15<g<19$, $\mu_{bhb}$ and $\sigma_{bhb}$ are relatively constant and they only deviate for stars at the very bright and faint end. This is seen in Figure \ref{fig:mcmc-g} which shows the results of the $\mu$ and $\sigma$ for the BHB and BS components as well as their fractions for several $g$-band intervals. At the faint magnitude range, the errors in the $grz$ colours are larger and thus strongly affect the BHB Gaussian distribution. \par
\citet{santucci+2015} showed that BHB stars identified spectroscopically have a shift in the $(g-r)$, being redder towards larger distances, due to an age difference of $\sim 2$ Gyr. Motivated by this, we repeated the previous exercise for two colour intervals, $-0.3<(g-r)<-0.175$ and $-0.175<(g-r)<-0.05$, and verified that, for $P(grz)$ given by Equation \ref{eq:Pgrz}, $\mu_{bhb}$ and $\sigma_{bhb}$ are consistent between the two intervals. 
\par
Based on these results, for the analysis described in the next sections, we complement the log-likelihood from Equation \ref{eq:Pgrz} by Gaussian priors on $\mu_{bhb}$ centred on 0.0005 with a width of 0.0005, and on $\sigma_{bhb}$, centred on 0.014 with a width of 0.001. We assume flat priors for $\mu_{bs}$, $\sigma_{bs}$, $\beta$, and the fractions $w_{i}$. Finally, in the next section, we included the criteria:
\begin{equation}
    14.5 < r < 21.5,
\label{eq:criteriarband}
\end{equation}
avoiding saturation problems of LSDR9 photometry and completeness in the bright and faint ends, respectively. In the right column of Figure \ref{fig:mcmc-g} we show how the free parameters vary with the Gaussian prior and including the criteria in Equation \ref{eq:criteriarband}. Now, the fraction of the BHB component decreases towards fainter magnitudes, as expected due to the density profile of the MW stellar halo \citep[see similar exercise in e.g.][]{deason+2011}. 

\subsection{BHB stars density}\label{sec:radial-density}
We now proceed and describe how we calculate the radial density profile for any given volume within the LSDR9 footprint. We use the BHB absolute magnitude relation from \citet{belokurov_koposov+2016}:
\begin{equation}
\begin{split}
        M_{g,bhb} = & 0.398 - 0.392(g-r)_0 + 2.729(g-r)_0^2 + \\
        & 29.1128(g-r)_0^3 + 113.569(g-r)_0^4,       
    \end{split}
    \label{eq:Mgbhb}
\end{equation}
convert to heliocentric distance using the distance-modulus equation, and calculate the Cartesian Galactocentric coordinates, $(x,y,z)$, assuming the Sun's position $(X_{\odot}, Y_{\odot}, Z_{\odot})=(-8.2, 0, 0.027)$ kpc \citep{mcmillan2017, Bennett&Bovy2019}. \par
The stellar halo of the MW is also known to be flattened \citep[e.g.][]{zinn+2014,iorio+2017, hernitschek+2018, wu+2022, yang+2022}, therefore we calculate the density profiles as a function of the flattened radius:
\begin{equation}
r_q = \sqrt{x^2 + y^2 + \left(\frac{z}{q(r)}\right)^2},
\end{equation}
with $q$ being the flattening parameter. Although it is consistently found that the inner halo is flattened while the outer halo is more spherical, there is no consensus on how flat the MW halo truly is and whether its flattening varies with Galactocentric radius or not. It is out of the scope of the current work to fit for the parameter $q$, thus we assume $q=0.77$ for $r<30$ kpc and $q=0.99$ otherwise. This is motivated by results found in \citet{hernitschek+2018} and is in good agreement with the range of values found in the literature. We also verified that the results discussed below hold for different values of $q$ that can be found in the literature. \par%
A necessary ingredient for spatial density estimation is not only the estimate of the number of stars but also the volume covered by the survey.
We can explicitly calculate the volume encompassed by a sample defined within a Galactocentric distance-module range, i.e.:
\begin{equation}\label{eq:distmodrange}
    a < g_0 - M_{g,bhb} < b,
\end{equation}
where $g_0$ is the extinction corrected apparent magnitude in the $g-$band and $M_{g,bhb}$ is given by Equation \ref{eq:Mgbhb}. 
For the same volume, we can also estimate $N_{bhb}$ and its error as described in Section \ref{sec:meth} thus providing us with the estimate of the BHB number density, $n_{bhb}$, and its error, $n_{bhb,er}$, within the volume constrained by Equation \ref{eq:distmodrange} and the LSDR9 footprint. \par
\subsection{Radial density profiles}\label{sec:rdp}
The methodology described in the previous sections allows us to construct the radial density profile for the full LSDR9 footprint as well as for any direction or patch in the sky. Throughout this work, we fit the number density profiles with a double power law model:
\begin{equation}\label{eq:dpl}
    n(r_{q}) =
\begin{cases}
    n_{\odot} r_{q}^{\alpha_{in}},& \text{if } r_{q} < r_{br}\\
    n_{\odot} r_{br}^{\alpha_{in}}(r_{q}-r_{br})^{\alpha_{out}},              & \text{otherwise}
\end{cases}
\end{equation}
where $n_{\odot}$ is the local density normalization, $r_{br}$ is the break radius, and $\alpha_{in}$/$\alpha_{out}$ is the inner/outer slope. \par
In the next section, we construct the density profiles calculating the BHB density and its error estimated as in Section \ref{sec:meth} at several bins in Galactocentric distance. Ideally, it is preferred to perform a fitting using all the available data points and not binned data, but our methodology relies on calculating the probabilistic fraction of BHB stars, and not identifying them individually. Therefore, we follow the approach of \citet{anders+2021} and estimate the free parameters of Equation \ref{eq:dpl} minimising the $\chi^2$ function using the binned profile and taking into account the error in $n_{bhb}$. We use flat priors for all the four free parameters: $n_{\odot}$, $r_{br}$, $\alpha_{in}$, and $\alpha_{out}$, and {\tt emcee} to sample from the posterior distributions. We take the median values as the parameter values, and the errors are from the 16th and 84th percentiles. 
\section{Results}\label{sec:results}
The top left panel in Figure \ref{fig:rprofile-ovd} shows the Mollweide projection in Galactic coordinates, ($\ell, b$) of the stars satisfying the data selection criteria described in Section \ref {sec:data}. We apply the methodology described in Section \ref{sec:methodology} to this data, and explore the Galactocentric radial density profile of the MW, as observed with LSDR9, and of regions with known identified substructures in Section \ref{sec:prof-ovd}, proceed to a higher angular resolution radial profile in Section \ref{sec:prof-heal}, and depict the anisotropy of the stellar halo in Section \ref{sec:aniso}. 
\begin{table*}
\centering
\begin{tabular}{ |p{1.4cm} p{2.5cm} p{1cm} p{1cm} p{2.5cm}| }
 Name & (l,b) [deg] & $d_{\odot}$ [kpc] &  $r_{br}$ [kpc] & $(\alpha_{in},$  $\alpha_{out})$ \\
 \hline
HAC-S & $ 30^{\circ} < l < 60^{\circ}$,\newline
        $ -45^{\circ} < b < -20^{\circ}$ & 10-20 &  $24.8^{+0.8}_{-1.0}$ & $(-3.0^{+0.1}_{-0.1},$$-5.3^{+0.2}_{-0.2})$\\[0.11cm]
HAC-N & $ 30^{\circ} < l < 60^{\circ}$, \newline
        $ 20^{\circ} < b < 45^{\circ}$ & 10-20 & $22.3^{+1.3}_{-1.2}$ & $(-2.8^{+0.2}_{-0.2},$$-5.7^{+0.3}_{-0.3})$\\[0.11cm]
VOD & $ 270^{\circ} < l < 330^{\circ}$, \newline
        $ -45^{\circ} < b < -20^{\circ}$ & 10-20 & $22.4^{+3.6}_{-3.4}$ & $(-2.8^{+0.4}_{-0.3},$$-4.2^{+0.2}_{-0.2})$\\[0.11cm]
Pisces & $(74^{\circ}, -47^{\circ})$ & 60-100 &  -- & $(-3.8^{+0.7}_{-2.5},$$-3.8^{+0.2}_{-0.3})$  \\[0.11cm]
\end{tabular}
\caption{The Galactic coordinates and heliocentric distance used in this work for Hecules-Aquila Cloud South/North \citep[HAC-S/N,][]{belokurov+2007, watkins+2009,sesar+2010,simion+2014}, Virgo Overdensity \citep[VOD,][]{juric+2008, bonaca+2012}, and Pisces \citep{nie+2015, belokurov+2019}. For the latter, we indicate the centre of a $20^{\circ}$ region satisfying the regions associated with Pisces as in \citet{belokurov+2019} and \citet{conroy+2021}. We also show the slopes and break radius, $r_{br}$, of a two-power law fit of these regions obtained in this work. For comparison, the average profile within the whole LSDR9 footprint has $r_{br} = 19.1^{+1.7}_{-1.8}$ kpc and $(\alpha_{in},$  $\alpha_{out})=(-2.9^{+0.2}_{-0.1},-4.5^{+0.1}_{-0.1})$}.
\label{table:ovds}
\end{table*}
\subsection{The radial density profile of the MW and known overdensities}\label{sec:prof-ovd}

\begin{figure*}
\centering
\includegraphics{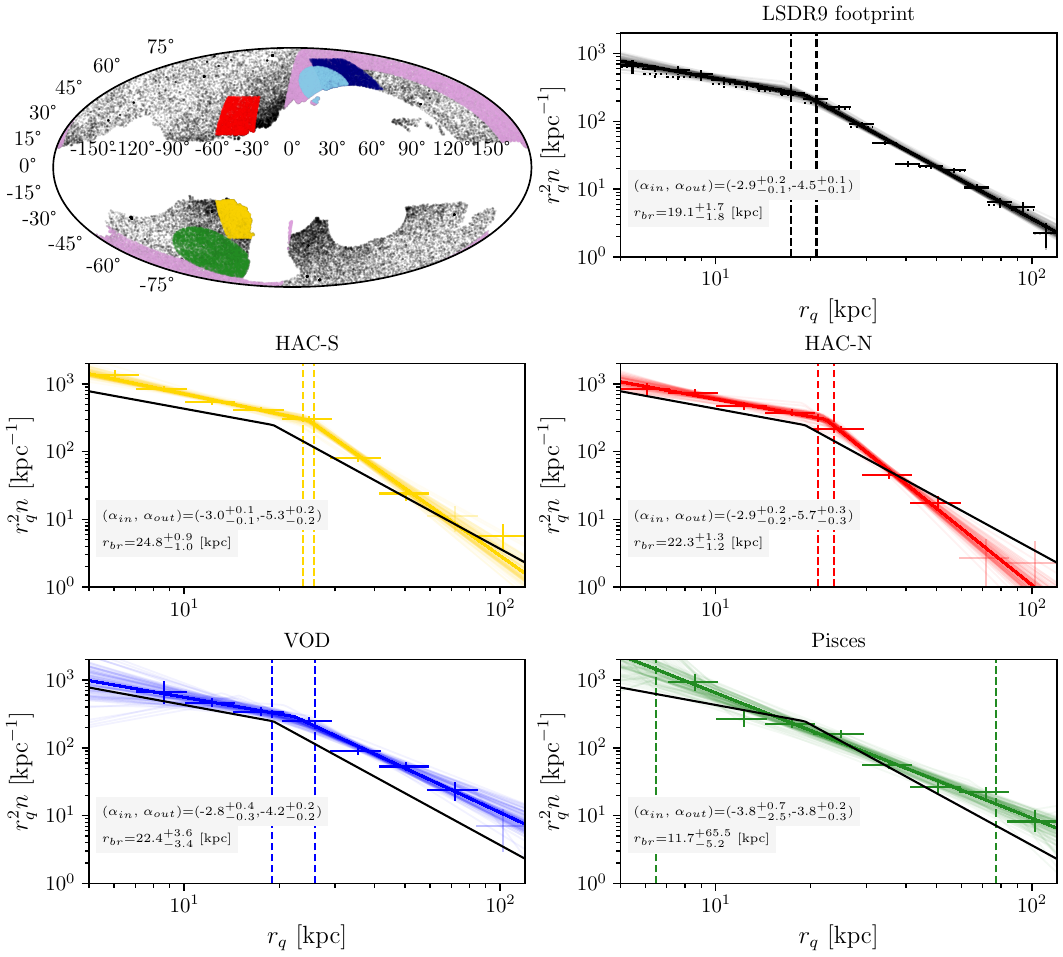}

    \caption{\textbf{Top left panel:} The Galactic Mollweide projection of LSDR9 stellar objects selected using Equation \ref{eq:cc}. The highlighted regions represent the location of known substructures: Sagittarius stream (pink), HAC-S (yellow), HAC-N (red), Pisces (green), VOD (blue) and OVO (cyan). \textbf{Top right panel:} The Galactocentric BHB stellar density profile of the whole LSDR9 footprint and without HAC-S, HAC-N and VOD regions as the solid and dotted markers, respectively. \textbf{Bottom panels:} The Galactocentric radial density profiles for the Galactic regions associated with HAC-N, HAC-S, VOD and Pisces. The crosses correspond to twenty and ten radial bins equally spaced in logarithm within $5 < r_{q}/ {\rm kpc} < 120$ for LSDR9 footprint and overdensities, respectively. We also show the resulting fit for 100 samples drawn from the posterior distribution as the solid lines.} 
    \label{fig:rprofile-ovd}
\end{figure*}
The top right panel in Figure \ref{fig:rprofile-ovd} shows the Galactocentric radial density profile for the whole LSDR9 survey sky coverage. The BHB stars average profile has a clear break at $r_{q} = 19.1^{+1.7}_{-1.8}$ kpc with inner and outer slope $(\alpha_{in}, \alpha_{out}) = (-2.85^{+0.17}_{-0.14},-4.55^{+0.11}_{-0.10})$. The location of the break and the inner and outer slopes are within the range of previously identified broken power law profiles in the MW halo, see for example table 6 in \citet{medina+2024} for a compilation of literature stellar halo density profiles. We defer the discussion and comparison with previous density profile estimations to Section \ref{sec:discuss}. \par
The MW stellar halo harbours several substructures/overdensities that could potentially alter its density profile shape and some of them have been linked to previous merger events. We identify the regions associated with Hercules Aquila Cloud North \& South (HAC-N \& HAC-S), Virgo Overdensity (VO) and Pisces overdensity, whose coordinates are given in Table \ref{table:ovds}, in the top panel of Figure \ref{fig:rprofile-ovd} as the coloured regions. We proceed and characterise the regions associated with the overdensities in the inner halo, $r_{q}< 30$ kpc, HAC-N/S and VOD, and outer halo, Pisces. The second and third rows in Figure \ref{fig:rprofile-ovd} show their radial density profile. The coloured markers correspond to the selected regions and the black solid line to the full LSDR9 profile, i.e. the average profile. \par
HAC-N, HAC-S and VOD have a broadly similar profile with a clear break radius, at the same position as the average profile, $r_{q} \approx 20$ kpc, and their slopes are within the errors as those estimated for the average sky. VOD shows an excess above the average density at all radii, while HAC-S and HAC-N only have excess within $r_{q} < 20$ kpc, i.e. within the distance range previously reported in the literature. Moreover, HAC-N/S have a steeper outer slope compared to the average profile and VOD. \par
As HAC-N, HAC-S, and VOD have been associated with GSE, which likely dominates the stellar halo mass content, and they show a clear signature of a broken profile, one could argue that the global profile is broken due to the presence of these regions. We thus mask these three regions, as defined in Table \ref{table:ovds}, and recalculate the global density profile. However, after removing HAC-N, HAC-S and VOD, the shape of the halo is still well represented by a broken power law, dotted markers in Figure \ref{fig:rprofile-ovd} top right panel, at odds with \citet{hernitschek+2017} whose methodology favours a single power law after removing these substructures. We stress, however, that we probe the halo from $5$ to $120$ kpc, while \citet{hernitschek+2017} only for $r_{q} > 20$ kpc. Nonetheless, the outer slope of our profile is in agreement with the slope these authors found.  \par
Among the regions with substructures, the Pisces region is the only one that shows no evidence of a break in its density profile. This is seen both in Figure \ref{fig:rprofile-ovd} but also in measurements of the inner and outer slopes, which are consistent within uncertainties, as well as by the fact that the break radius is not well constrained. Pisces overdensity was identified at large radii, and here we detect excess in its profile at $r_{q} > 50$ kpc. At its direction, there is also an overdensity at $r_{q}\approx 8$ kpc not previously reported. Furthermore, the density profile in its direction is less steep than the average outer stellar halo density profile. \par %

The behaviour of the density profile in four different regions hints at an anisotropic density profile of the MW stellar halo which could be ``direction-dependent" rather than being described by a single density profile stratified on simple surfaces like ellipsoids. Motivated by this, we proceed to study the radial density profile for several regions in the sky as seen in the Galactocentric angular coordinate system.
\subsection{The radial density profile at distinct Galactocentric angular coordinates}\label{sec:prof-heal}

In this subsection, we show the radial density profile in several Galactocentric directions. We convert from Galactocentric Cartesian $(x,y,z)$ to spherical coordinates and define the spherical angles $\Theta_{gc}$ and $\Phi_{gc}$ as the Galactocentric longitude and latitudes, respectively:
%
\begin{align}
\Theta_{gc} =  \arctan{\frac{z}{\sqrt{x^2+y^2}}},  &&  \Phi_{gc} = \atantwo(y,x)
\end{align}

\begin{figure*}
\centering

\includegraphics{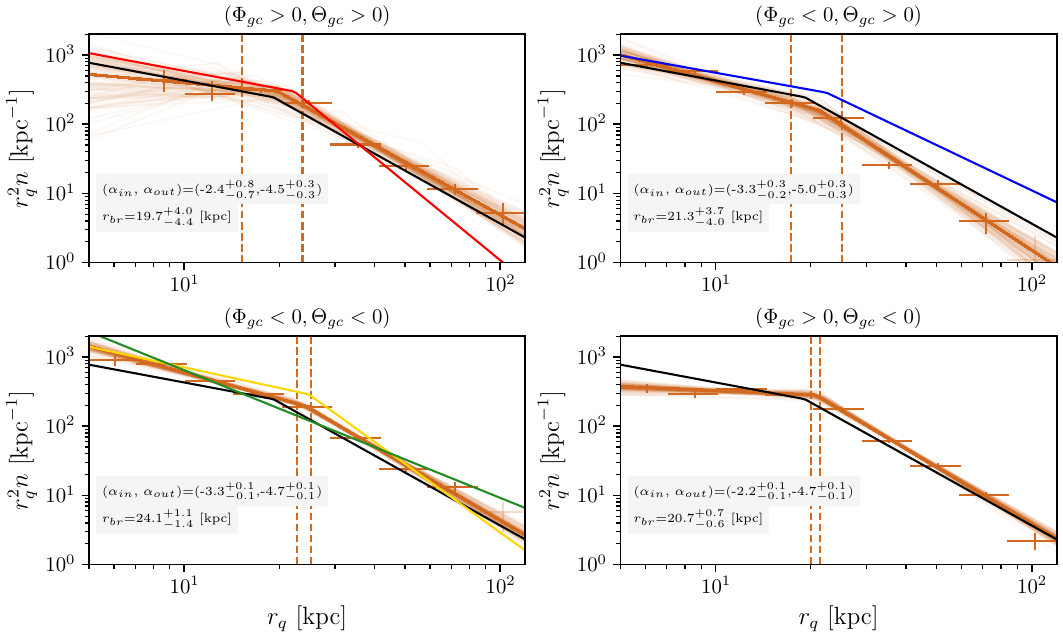}
    \caption{The Galactocentric radial density profiles for the four quadrants of the Galactocentric projected sky. The crosses correspond to ten radial bins equally spaced in logarithm within $5 < r_{q}/ {\rm kpc} < 120$. We also show the resulting fit for 100 samples drawn from the posterior distribution as the brown solid lines. The histogram corresponds to the posterior distribution of the break radius, $r_{br}$, and the dashed vertical lines are the corresponding 16th and 84th percentiles. The density profile of the full LSDR9 footprint, HAC-N, HAC-S, VOD and Pisces are shown as the black, red, yellow, blue, and green lines, respectively.}
    \label{fig:rprofile-quad}
\end{figure*}
We first show the density profile for the four quadrants of the Galactocentric projected sky in Figure \ref{fig:rprofile-quad}. We do not remove the substructures located within each quadrant. While the four quadrants have signatures of a break and similar outer slopes, within the errors, they differ in the steepness of the inner slopes, $\alpha_{in}$. For instance, the profile of the second ($\Phi_{gc} >0 , \Theta_{gc}>0$) and fourth ($\Phi_{gc} >0 , \Theta_{gc}>0$) quadrants are not as steep as the first ($\Phi_{gc} < 0 , \Theta_{gc}>0$) and third ($\Phi_{gc} <0 , \Theta_{gc}<0$) ones. \par
There are also noticeable differences when comparing the profile of the regions associated with substructures to the Galactocentric quadrants in which they are located. HAC-N shows a distinct inner density profile, within the break radius, compared to the second quadrant, where it is located. This is shown in the left panel of Figure \ref{fig:rprofile-quad}, where the brown markers represent the radial density for the quadrant and the red line is the best fit for HAC-N, as in the previous section. This quadrant's profile closely follows the average profile (solid black line) in the inner parts, even without masking HAC-N when measuring it, in contrast to HAC-N. This reinforces the presence of HAC-N as a genuine overdensity, as this region neither follows the global profile nor that of the quadrant it is located. \par
The third quadrant, shown in the bottom left of Figure \ref{fig:rprofile-quad}, includes two overdensities regions: HAC-S and Pisces, whose profiles are shown as the yellow and green lines respectively. While in the inner parts, the quadrant's profile shows overall overdensity compared to the average one, its outskirts follows closely the average profile, in contrast to the Pisces region. \par
The first quadrant, top right panel of Figure \ref{fig:rprofile-quad}, harbours the region associated with VOD. Its inner part follows closely the average profile, in contrast to the VOD profile (blue solid line), whereas the outer part also shows some overall underdensity. Finally, the fourth quadrant, which harbours the LMC and SMC but no known overdensity, differs only in the inner parts to the average profile having a less steep inner slope. \par
The differences and similarities between each quadrant, together with the regions associated with each overdensity, point to a directional dependence on the radial density profile which is generally not taken into account in previous studies. We investigate further the apparent asymmetry, and pixelized the projected Galactocentric sky using the Python package {\tt healpy} \citep{healpix, healpy}. We use $nside=2$ which divides the sky into 48 patches (healpixels) each with an equal projected area of 859.44 ${\rm deg}^2$. Note that the LSDR9 footprint does not cover all the patches and a fraction of them are not fully covered. Nevertheless, the density for each region defined by a healpixel takes into account the LSDR9 footprint.\par
We focus on the regions which contain known overdensities with a coherent coverage in Galactocentric radius and show their radial density profile in Figure \ref{fig:rprofile-hp} (see Appendix \ref{app:hp-profile} for the profile of all patches). We identify the patch location in the Mollweide projection of the Galactocentric coordinates in the top right corner of each panel and its ``nested ID'' number\footnote{The {\tt healpy} package orders the healpixels either in a nested or ring scheme.}. The purple crosses are the average density profile for the highlighted patch and the solid black line corresponds to the best fit for the LSDR9 footprint. For each region, we fit a double power law profile and show the interquartile of the posterior distribution of the break radius as the dashed vertical lines. We also explicitly highlight the locations associated with known substructures: HAC-N (red), HAC-S (yellow), Pisces (green), VOD/Outer Virgo Overdensity (OVO) (blue) and Sgr (pink). The less transparent the colour, the lower the contribution of the given substructure in a given pixel on the sky\footnote{The contribution of the substructure to a pixel is computed as a fraction of the pixel area that intersects with the substructure area given in Table \ref{table:ovds}.}. For Sgr we use the trail and leading arms distance-relation from \citet{hernitschek+2017} (their table 4 and 5). We define OVO as a $20^{\circ}$ region centred in right ascension and declination, ${\rm R.A.} = 207^{\circ}, {\rm Decl.}= -7^{\circ}$, respectively \citep{sesar+2017-ovo}.  \par
\begin{figure*}
\centering
\includegraphics[width=\linewidth]{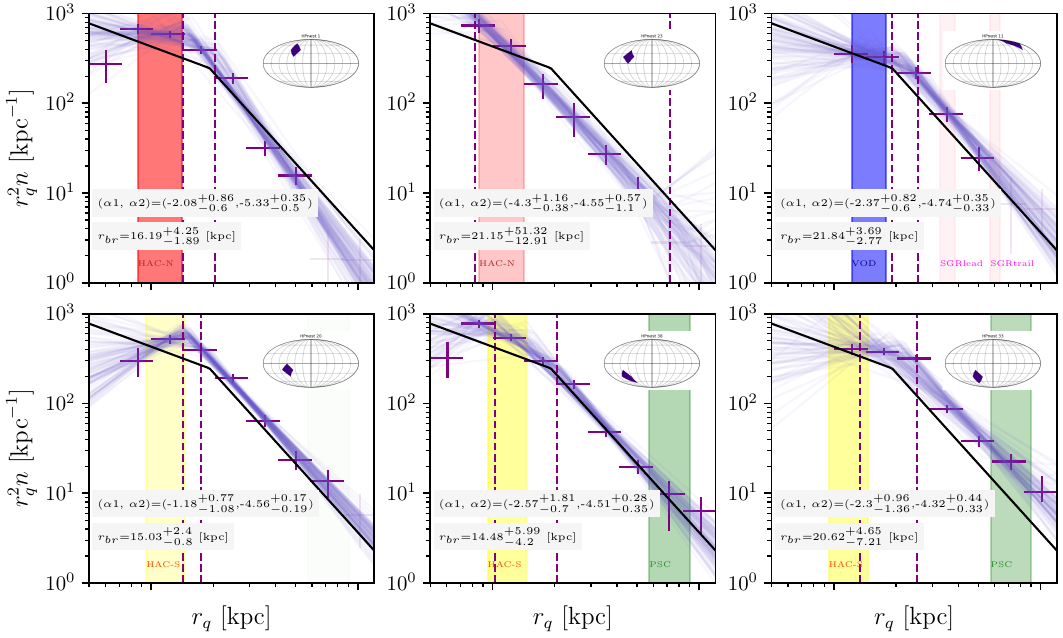}
    \caption{The radial density profile for several patches in the Galactocentric sky covering an area of $859.44\,{\rm deg}^2$. The purple crosses correspond to the radial densities of the highlighted region shown in the Mollweide map of each panel. The purple solid lines correspond to the double power law fit using 100 samples drawn from the parameters' posterior distribution, and the dashed vertical lines correspond to the 16th and 84th percentiles of the $r_{br}$ posterior distribution. The black solid line is the best fit performed on the whole LSDR9 footprint. 
    Each panel also shows the contribution of known substructures in a given region: HAC-N (red), VOD (blue), HAC-S (yellow), Pisces (green) and Sgr stream (pink). The darker the highlighted region is, the stronger the contribution of the substructure on the highlighted Galactocentric region.}
    \label{fig:rprofile-hp}
\end{figure*}
Figure \ref{fig:rprofile-hp} shows the strength of the methodology developed in Section \ref{sec:meth}. By assigning a probabilistic BHB number for a given direction, we have a more detailed stellar halo profile compared to previous studies and can study any angular difference between the profiles. A similar approach was made by \citet{hernitschek+2018} as they probed the halo in several slices in $(l,b)$. However, they sliced in patches which cover an area in the sky almost twice as large as the ones depicted here, removed HAC and VOD overdensities, and excluded the inner parts of the Galaxy, $r_{q} < 20$ kpc. We now detail the features in those regions which include the known overdensities.  \par
The top left and middle panels in Figure \ref{fig:rprofile-hp} show the density profile of two patches which include the HAC-N overdensity, as defined by the coordinates given in Table \ref{table:ovds}. The former contains most of the overdensity, as highlighted by the darker red band compared to the latter, and its profile shows a clear break radius at $r<20$ kpc. At larger radii, the profile has a steeper decrease compared to the average LSDR9, but we note that the last two radial bins have a large uncertainty. On the other hand, the radial profile of the patch shown in the middle panel is consistent with a single power law and also shows an overall underdensity at radii greater than 20 kpc.\par
The top right panel in Figure \ref{fig:rprofile-hp} shows one of the sky patches that includes VOD. In this direction, the Sgr leading and trailing arms are expected to contribute at $r_{q} \approx 35$ kpc and $r_{q} \approx 60$ kpc, respectively, \citep[as in][]{hernitschek+2017}. Although we have removed stars associated with Sgr (Section \ref{sec:contamination}), there is still a slight overdensity at the distances associated with Sgr's leading arm. Moreover, the overdensity associated with VOD seems to be at a slightly larger distance than what was previously reported, and closer to the break radius, i.e. at $r_{q} \approx$ 20 kpc. \par
The bottom row in Figure \ref{fig:rprofile-hp} shows the density profile in three patches of the Galactocentric sky which contain HAC-S and Pisces overdensities. In this coordinate system, they overlap in angular space, a feature which is missed in the Galactic coordinate system (see the top left panel in Figure \ref{fig:rprofile-ovd}). One can visualise it as being at the centre of the MW, looking towards a direction in the sky and seeing through two overdensities located at different radii. Firstly, the inner parts, located within the break radius, consistently show an overdensity associated with HAC-S. Secondly, the overdensity associated with Pisces seems to be mainly located in the patch centred at $\Phi_{gc} \approx 67.5$ deg and $\Theta_{gc}\approx -41.8$ {deg}\footnote{Nested healpix number 33, for $nside = 2$.}, and shown in the bottom right panel. This panel shows an overdensity through all radii, due to HAC-S and Pisces, and its outer slope is as steep as the average halo. These facts have implications on the interpretation of the possible origin scenarios of Pisces, as we will discuss in Section \ref{sec:discuss}. \par

\subsection{The anisotropy of the halo density profile}\label{sec:aniso}
\begin{figure}
\centering
\includegraphics[width=\linewidth]{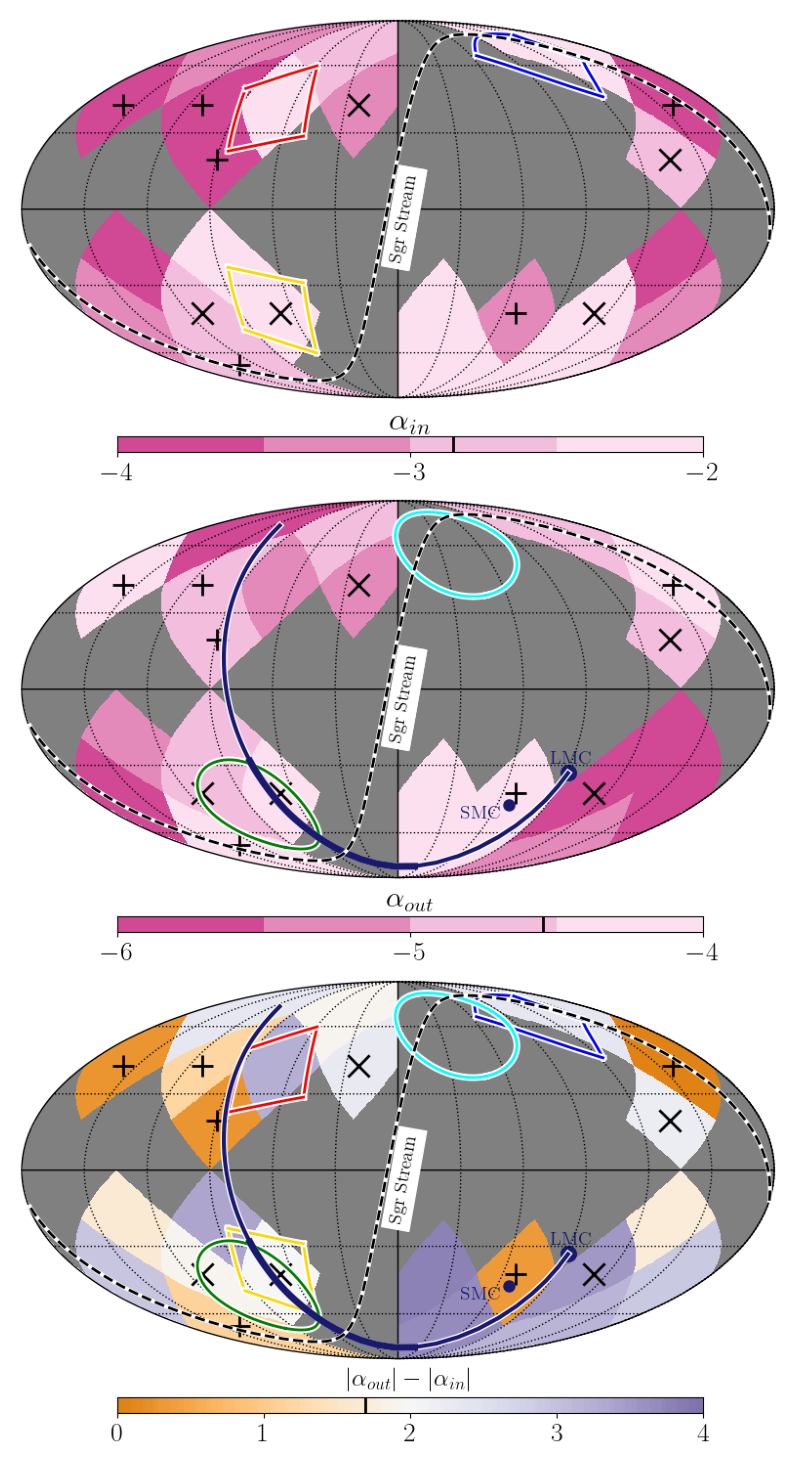}
    \caption{Mollweide projection of the Galactocentric coordinates $(\Phi_{gc}, \Theta_{gc})$. The projection is divided into a grid of 48 pixels ($nside=2$) with equal area. The blue, red, yellow and orange regions delineate the regions associated with VOD, HAC-N, HAC-S and Pisces at their typical distances.  \textbf{Top panel:} The inner slope, $\alpha_{in}$, measured at $r_{q} = 15$ kpc, of the two power law model. \textbf{Middle panel:} The outer slope, $\alpha_{out}$, measured at $r_{q} = 50$ kpc, of the two power law model. \textbf{Bottom panel:} The difference between  $\alpha_{in}$ and $\alpha_{out}$. These panels highlight the anisotropy of the MW stellar halo profile, where several patches of the sky have distinct slopes compared to the average halo. We only show patches with more than four density points in the radial binning. The crosses and ``x" indicate the regions with $|\alpha_{out}|-|\alpha_{in}|<1.5$, and $|\alpha_{out}|-|\alpha_{in}|>1.5$ with error in the break radius greater than 25\%, respectively.}
    \label{fig:moll-feat}
\end{figure}

The anisotropy of the halo density profile is depicted in Figure \ref{fig:moll-feat}. The top and middle panels show the inner and outer slopes, respectively, of the double power law model fit for regions shown in Figure \ref{fig:panel}. We only show the regions with more than four radial density bins available. The crosses indicate the patches where $|\alpha_{out}|-|\alpha_{in}|<1.5$, the orange-coloured pixels in the bottom panel. These patches do not show strong evidence for a broken power law profile, or the least their difference is lower than what is observed in the global profile.  \par
The inner parts of the halo show significant differences in the profile across the Galactocentric sky, as indicated by $\alpha_{in}$. For instance, the regions associated with HAC-N, HAC-S and VOD have a flatter inner profile compared to the average LSDR9 profile, as indicated by the lighter-coloured pixels shown in the top panel, in contrast to most of the patches, except for regions in the fourth quadrant. \par
The regions marked with ``x" indicate where $|\alpha_{out}|-|\alpha_{in}|>1.5$, white- to purple-coloured pixels in the bottom panel, but the error in the break radius is larger than 25\%. Interestingly, this is observed in the regions where HAC-S and Pisces regions overlap (as shown in Figure \ref{fig:rprofile-hp}), whereas the patch which includes only Pisces does not show evidence of a broken profile. The other three patches with significant differences between inner and outer profiles, but with $r_{br}$ poorly constrained, are either close to the Galactocentric plane, or near Sgr stream or LMC stellar halo contaminants.\par%
Recently, \citet{medina+2024} showed how the density profile varies for four small patches, area $\lesssim 80\, {\deg}^2$, in the sky. They also highlighted the dependence of the halo density profile on the regions probed by a survey and the stellar tracer used. The results presented here do confirm significant differences in the stratified stellar halo, for rather larger sky patches, when compared to the global average. This has not been shown previously, as most studies either focused on a particular patch of the sky or did not probe the halo with a similar Galactocentric radial extent as done here. %

\section{Discussion}\label{sec:discuss}
In this current work, we found that the average density profile of the MW halo, as traced by the BHB stars, is described by a double power law (Figure \ref{fig:rprofile-ovd}, top right panel) corroborating with a plethora of similar studies. \citet{medina+2024} argues that the apparent lack of consensus on the exact profile shape, e.g. either single- or double power-law, can be understood by the tracer being used, the magnitude range available, and the spatial selection used, such as the minimum/maximum Galactocentric radius in the density profile construction. In addition to that, and most importantly, the discrepancy found in the literature is also caused by the heavily anisotropic halo shown here. \par%
Overall, the studies which find evidence for a single power law profile do not cover the inner parts of the Galaxy, i.e. $r_{gc} \lesssim 20$ kpc. Those which include the inner halo, in general, do find evidence for a double-power law profile with outer slopes with an overall agreement with the slope of the single-power law studies {(see e.g. recent discussion in \citealt{yu+2024}) and \citealt{medina+2024} and their compiled list of previous works on the MW halo density profile}). \par
In Section \ref{sec:results}, we have shown evidence for a broken power law (BPL) density profile for the BHB stars in LSDR9 with a break radius at 20 kpc. The BPL profile is also observed when looking at the directions of HAC-N/S and VOD overdensities and in some patches in the Galactocentric sky. Moreover, we do find significant anisotropy in the inner and outer halo as several patches have distinct slopes, see Figures \ref{fig:moll-feat} and \ref{fig:alphas}, where the latter also compares with slopes found in previous works. Moreover, some regions have a profile consistent with a single power law, e.g. in the direction of the Pisces overdensity.\par
We proceed and discuss how our results can shed light on the understanding of the structure of the stellar halo. We separate the discussion into two sections; in Section \ref{sec:inner} we discuss the inner halo which is mainly associated with the GSE debris, $r_{gc} < 20$ kpc, and in Section \ref{sec:outer} we focus on the outer halo which hosts imprints of the LMC passage.
\subsection{The inner halo and the GSE debris}\label{sec:inner}
\citet{deason+2013} made the connection between the presence of a broken profile and the merger history of the MW using \citet{bullockjohnston2005} simulations. The break radius corresponds to the apocenter pile-up of the debris from a past massive merger. The inner part of the halo, within the break radius, is dominated by the debris of the main merger and thus has a less steep density profile compared to the outer parts, dominated by minor accretion events. As they observed this in the MW, they concluded the MW had gone through a main accretion event in its early stages. This was yet another strong evidence among others \citep[e.g.][]{chiba&beers2000,gilmore2002, meza2005}, of the GSE merger event later confirmed with Gaia data \citep{belo2018, helmi2018, haywood}.\par 
More recently and thanks to \textit{Gaia} and large spectroscopic surveys, the halo density profile is also built with halo star candidates selected based on stellar chemistry and kinematics \citep{han+2022, yang+2022, lane+2023}. This methodology relies on the fact that the GSE debris is the major contributor to the stellar halo mass \citep[e.g.][]{naidu+2020}. Intriguingly, this approach has also provided evidence for a doubly broken halo, i.e. a three-power law model, when selecting GSE star candidates chemically \citep{han+2022} or with a metallicity-kinematic decomposition \citep{yang+2022}. Each break radii is associated with {the apocenters of GSE orbit during its merger}\citep{naidu+2021_nbody}.  \par
Besides that, the GSE debris has been associated with HAC-N/S and VOD substructures. They have been shown to share similar dynamics \citet{simion+2019} and chemodynamical \citep{perottoni+2022} signatures as the GSE near the Solar neighbourhood. {In Section \ref{sec:prof-ovd}, the radial density profile of these regions is shown for the first time for a large distance extension, $5 < r_{q}/{\rm kpc} < 120$, which allows showing clear evidence for the presence of a broken radial profile. These regions share a similar $r_{br}$ ranging from $\sim 22-24$ {kpc}, similar inner and outer slopes, and a clear signature of the overdensity in the inner parts, where these substructures have been identified. The location of their break radius is also $\sim 1-2 \sigma$ larger than the global average. The break radius of the halo profile has been associated with the apocenter of the GSE dwarf orbit \citep[e.g.][]{naidu+2021, han+2022}. Thus, the difference between the break radius of the overdensities and the global average is unexpected in the scenario where the GSE dominates the inner halo and these overdensities are part of its debris.  }
{We also show in Figure \ref{fig:rprofile-quad} that these overdensities' profiles are also consistent with the Galactocentric quadrants where they are located. In Section \ref{sec:prof-heal} we further explored these regions in smaller patches in the Galactocentric sky reinforcing the signature of the overdensities. } \par
The density profile for these specific overdensity regions associated with the GSE merger serves as a constraint to idealised simulations that mimic this merger event. For instance, \citet{naidu+2021_nbody} presented an N-body model, without gas and star formation, which reproduces several properties of the observed GSE debris, including HAC-S and VOD overdensities, but does not reproduce the substructure in the HAC-N region. The latter is also an important constraint, and its overdensity is also observed in previous studies \citep[e.g.][]{huang+2022}, due to its similarity with HAC-S and VOD. In simulations with gas and star formation, either tailored \citep{amarante+2022} or in fully cosmological context \citep{orkney+2023},
the debris of GSE-like mergers fully phase mixes in the inner parts after a few Gyr and do not show evidence for clear overdensities like HAC-N/S and VOD. \par
To complicate matters further, \citet{orkney+2023} has also shown evidence that MW-like galaxies in the Auriga simulations \citep{grand+2017} can have stellar halos with high radial velocity anisotropy, i.e. like the GSE component in the MW, being built with up to three main dwarf satellites of similar mass. These accreted stars have enough similarity in chemistry and dynamical space making it almost impossible to distinguish them apart solely on [Fe/H], [$\alpha$/Fe], and with dynamics. {Moreover, \citet{donlon-newberg+2023} discusses the possibility of multiple dwarfs being able to explain the chemodynamics of the inner halo, where the GSE is claimed to be the main contributor. We note, however, that in situ contamination can contaminate halo-selected samples \citep{dasilva+2023}. }
\par
It remains to be seen, whether the presence of the HAC-N, HAC-S and VOD can be indeed a feature of a unique merger event which happened 10 Gyrs ago, or if any of these overdensities are indeed signatures of a more recent merger as argued by \citet{donlon+2020, donlon+2023}.
\subsection{The outer halo and the LMC wake}\label{sec:outer}
\begin{figure}
\centering

\includegraphics[width=\linewidth]{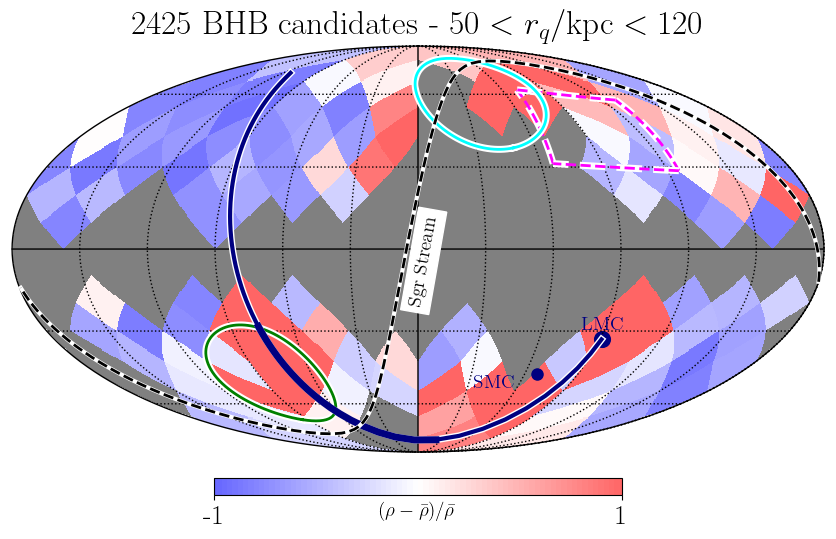}

    \caption{Density contrast of MW BHB stars in the stellar halo at $50 < r_{q}/{\rm kpc} < 120$ seen in Galactocentric coordinates $(\phi_{gc}, \theta_{gc})$. The Mollweide projection is divided into a grid of 192 pixels ($nside=4$) with equal area. We delineate the approximate Galactocentric regions associated with OVO, Pisces, and Northern overdensity as the cyan, green, and magenta lines, respectively. The present-day position of the LMC and SMC are shown as the purple scatter points. LMC's past trajectory (see \citealt{vasiliev2023} for details) and Sgr stream are represented by the solid purple and dashed black lines, respectively.}
    \label{fig:moll-contrast}
\end{figure}

To have a general picture of the outer halo, we divided the Galactocentric sphere into 192 equal-sized area pixels, using {\tt healpy} Python package with HEALPIX pixelisation $nside=4$, and calculated the halo density at $50 < r_{q}/{\rm kpc} < 120$ using the methodology described in Section \ref{sec:meth}. We then calculate the density contrast and show it in Figure \ref{fig:moll-contrast}. The LMC past trajectory is shown as the purple line \citep{vasiliev2023}\footnote{We made use of the LMC-MW interaction example of the {\tt AGAMA} python library \citep{agama}.}, with the thicker part corresponding to the orbit within the distance range shown in the map. \par
Although we removed stars within 12.5 degrees from the LMC centre, there is still an overdensity seen in the pixels surrounding LMC's current position. These are likely contaminants of LMC star members, as the LMC stellar halo can extend up to 30 degrees from the LMC as seen with BHB stars \citep{belokurov_koposov+2016} and RR Lyrae \citep{petersen_lmcstripped}. \par 
It is also evident from Figure \ref{fig:moll-contrast} {the presence of several regions} with a density contrast close to $-1$. Due to the steep density profile at these distances, $\alpha_{out}=-4.55$, the average density is dominated by the regions with overdensities. In fact, $\sim 55\%$ of the BHB stars, as estimated with our methodology, is within $\sim 20\%$ of the pixels. If the BHB stars followed a purely Poisson distribution, the same amount of stars would be within $\sim 40\%$ of the pixels. This reinforces the highly anisotropic stellar halo of the MW being dominated by a few patches with overdensities. \par
The Pisces overdensity, as highlighted approximately by the green circle in Figure \ref{fig:moll-contrast}, has been recently associated with the transient dark matter wake caused by the LMC passage. Its signature has been associated with an overdensity of RR Lyrae \citep{belokurov+2019}, K-giant stars \citep{conroy+2021}, {main-sequence stars \citep{suzuki+2024}}, and is now also observed with the probabilistic BHB star mapping developed in this work. This transient response of MW's (dark) halo to the LMC is akin to the Chandrasekhar dynamical friction wake \citep{chandrasekhar1943-dynfric} and traces the orbital path of the LMC in the halo at large distances \citep[e.g.][]{garavito-camargo+2019}. The shape of this transient wake signal should also be proportional to the mass and velocity of the LMC (and the DM density of the ambient medium). {Moreover, \citet{rozier+2022} showed that the detailed geometry of the local wake is also sensitive to the underlying anisotropy, tangential or radial, of the tracers.} Our BHB stars map can thus be used to measure the extent of the transient wake response. To this end, we study the BHB star density variations along the sky patches crossed by the LMC's past orbit from \citet{vasiliev2023}) for $50 < r/{\rm kpc} <120$.\par
\begin{figure}
\centering
\includegraphics[width=\linewidth]{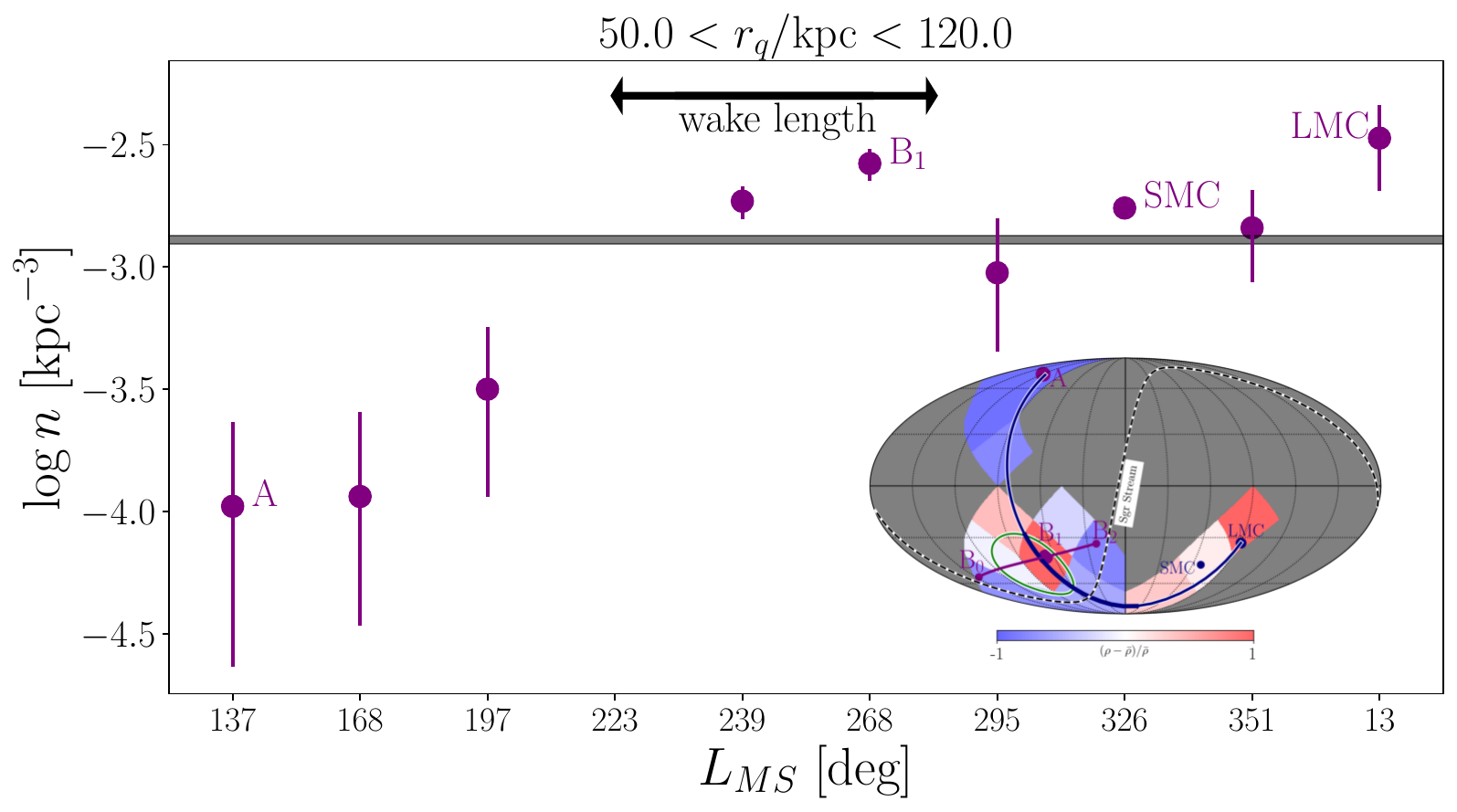}
\includegraphics[width=\linewidth]{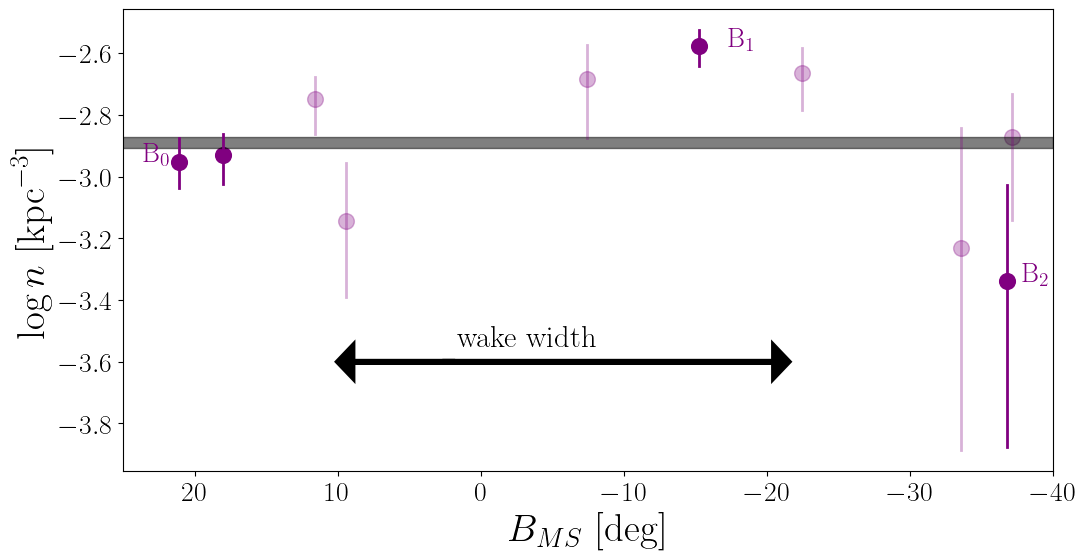}
\caption{The BHB star density of the sky patches crossed by the LMC's past orbit \citep{vasiliev2023} as a function of the $(L_{MS}, B_{MS})$ coordinates \citep{nidever+2008}. Each patch has an equal projected area of 859.44 ${\rm deg}^2$ ({\tt healpy} using $nside=2$) and the density is calculated within $50 <r/{\rm kpc} < 120$. The 16th-84th percentiles of the average BHB density within this distance is shown by the shaded green line. {\textbf{Top panel:}} The density profile along LMC's orbit from point $A$ to the pixel containing LMC.  At $B_1$, the pixel dominated by Pisces overdensity shows a clear signature of the overdensity in contrast to regions further away from it at $L_{MS}<230^{\circ}$. {\bf Bottom panel:} The density profile as a function of $B_{MS}$ from point $B_0$ to $B_1$. At these large distances, and due to the steepness of the profile, the average density is dominated by the regions with the overdensities as illustrated in both panels. {The lighter purple points correspond to the density at smaller patches ({\tt healpy} using $nside=4$) across the LMC's orbit.}} 
\label{fig:dens-lmc-orbit}
\end{figure}

The top and bottom panels of Figure \ref{fig:dens-lmc-orbit} show how the density varies along the Magellanic Stream Coordinates\footnote{In this reference system, LMC is located at $L_{MS} \approx 0^{\circ}$.} \citep{nidever+2008}. For this exercise, we coarsened the angular resolution, compared to Figure \ref{fig:moll-contrast}, to reduce the Poisson noise. Despite the relatively large patches, it clearly shows the density peak at the Pisces region, point $B_1$, and how it decreases going further away in $L_{MS}$ and $B_{MS}$.
From Figure \ref{fig:dens-lmc-orbit}, we can {estimate the upper limit of the Pisces overdensity extension to be} 60 degrees along and {30} degrees across the LMC’s orbit, which would correspond to a wake width of $\sim 32$ kpc at $\sim 70$ kpc. Future density profile maps along the orbital path of the LMC, as those shown in Figure \ref{fig:dens-lmc-orbit}, to characterise the transient wake signal will help further constrain models of the MW-LMC interaction and provide an important test for the nature of dark matter on Galactic scales.\par
We note in passing that the Pisces overdensity region may also overlap with the Magellanic Stellar Stream (MSS, \citealt{chandra+2023}) and {may also contain GSE debris.}
{As discussed previously, VOD and HAC are associated with the GSE merger, and the former shows an overall overdensity at larger radii (see Figure \ref{fig:rprofile-ovd} and \citealt{vivas+2001}). As HAC-S and Pisces overlap in angular position on the Galactocentric coordinate system (see e.g. Figure \ref{fig:rprofile-hp} bottom right panel), it is intuitive to expect some contribution to the GSE debris on this region, potentially biasing the effect of the wake signal. Kinematics together with chemical abundance information could help disentangle the MSS from the wake of the LMC and the contribution, if any, of the GSE debris to Pisces overdensity.} \par
Simulations of the MW-LMC interaction also predict the barycenter shift of the MW-LMC system \citep[e.g.][]{gomez+2015,garavito-camargo+2019}, i.e. the gravitational collective response or ``global wake". This effect has been observed kinematically \citep{erkal+2021,petersen_penarrubia2021, yakib+2024, chandra+2024} and the density dipole associated with the collective response \citep{weinberg1998,laporte+2018a, petersen_wakesim} should also, in principle, be observed in star counts  \citep{garavito-camargo+2019}. Recently, \citet{conroy+2021} have reported a possible detection of this signal using K giant stars. Given that BHBs come in greater numbers, we attempt to measure the signal from the dipole in star counts. \par
We show in Figure \ref{fig:moll-contrast} the approximate region in the Galactocentric sky of the region defined by \citet{conroy+2021}, magenta dashed-line, which we refer to as the Northern overdensity. The BHB density contrast map shows no apparent significant overdensity in this region, but there is an overdensity just to the north of the Northern overdensity. Part of this region is located in the region associated with OVO, indicated by the cyan circle, and relatively close to the Sgr Stream, the latter being mostly removed. Nonetheless, this may be the remains of Sgr at very large distances \citep[e.g][]{sesar+2017}.\par
\begin{figure}
\centering
\includegraphics[width=\linewidth]{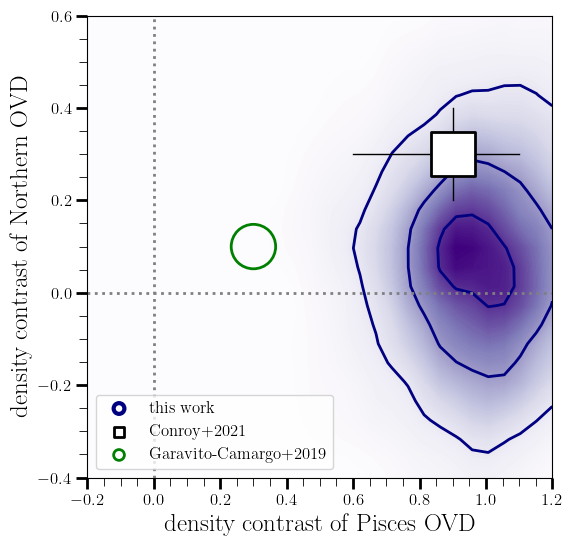}
\caption{The BHB density contrast at $50 < r_{q}/{\rm kpc} < 120$ in the Pisces overdensity and the Northen overdensity regions, which are associated with the local wake and the collective response to the LMC orbit respectively. The purple contours to the shaded region correspond to the 16th, 50th and 84th confidence intervals calculated from the BHB fraction posteriors (see text for details). This result can not confirm the collective response of the MW stellar halo to the LMC passage. For comparison, we show previous measurements using K giant stars \citep{conroy+2021} and predictions from MW-LMC interaction models \citep{garavito-camargo+2019}.}
\label{fig:contrast}
\end{figure}

Figure \ref{fig:contrast} shows the total density contrast measured in the Pisces overdensity region and the Northern overdensity which are associated with the transient and global wake respectively. The purple contours show the interquartile of the density contrast measured obtained from the posteriors of the density estimation. While Pisces overdensity shows a clear signature where the density contrast is $>0.6$, the signal measured in the Northern overdensity has a large error and is consistent with being negligible. {While the dipole signal in stellar density is observed in N-body simulations \citep[e.g.][]{garavito-camargo+2019},} the current BHB star data falls short of making a statistically meaningful measurement of this signal. This may need to await tracers that come in larger numbers in order to detect this signal, e.g. Main Sequence Turn Off stars. \par
We also show the results predicted from \citet{garavito-camargo+2019} simulations which have significantly lower signatures in the density map. This can be caused by the fact that their stellar halo started as a smooth component and in the MW the stellar halo is heterogeneous and thus the LMC can enhance some previously existing overdensity(ies). Indeed as shown in this study and others \citep[e.g.][]{bell+2008}, the stellar halo beyond 50 kpc is not smooth but inhabited by substructure which due to the long orbital timescales have not had time to fully phase mix over a Hubble time \citep[e.g.][]{johnston+2008}.\par

\section{Conclusion}\label{sec:conclusion}
We used LSDR9 photometric survey to study the density profile of the MW stellar halo. We developed a novel approach to modelling the $g-r$ and $r-z$ colour distribution allowing us to compute the fraction of BHB stars at any given direction. With this approach, we can robustly estimate the errors of the densities. We summarise our main results when probing the stellar halo within $5 < r_{q}/{\rm kpc} < 120$:

\begin{itemize}
    \item The global stellar halo shows a broken power law density profile with inner and outer slopes, and break radius, $\alpha_{in}=-2.45^{+0.17}_{-0.14}$, $\alpha_{in}=-4.55^{+0.11}_{-0.1}$, $r_{br}=19.15^{+1.7}_{-1.8}$ kpc, respectively;
    \item HAC-S, HAC-N and VOD Galactic structures show a significant break in stellar density profiles at $r_{q}\approx20$ kpc, with compatible density inner slopes (Figure \ref{fig:rprofile-ovd}). The presence of these three overdensities regions remains to be seen in tailored GSE-like merger models;
    \item The Pisces overdensity region is the only region where we see a density profile consistent with a single power law, in contrast to the previously mentioned overdensity regions;
    \item Even after excluding regions containing major overdensities we observe a density profile with a break which seems to indicate that the GSE debris is present in most directions in the Galaxy {or alternatively another structure is present in the Galaxy that also has a broken radial density profile with similar break radius}; 
    \item We find strong evidence for an anisotropic stellar halo, indicated by significant differences in the steepness of the density profile when looking at different Galactocentric patches (Figure \ref{fig:moll-feat}) and by $\sim 20\%$ of the sky containing $\sim 55\%$ of BHB stars at $r>50$ kpc (Section \ref{sec:outer});
    \item The Pisces overdensity spans 60 degrees along and {30} degrees across the LMC's orbital path in the sky (Figure \ref{fig:dens-lmc-orbit}); 
    \item If thought to be associated with the transient wake response of the MW to the LMC, then the Pisces overdensity extension would correspond to a wake width of $\sim 32$ kpc at $\sim 70$ kpc (Section \ref{sec:outer}). Such a measurement may help set constraints on the mass and models of the MW-LMC system and also on the nature of dark matter;
    \item While we confirm the overdensity in the Pisces region, associated with the local wake induced by the LMC passage, our results do not show a statistically significant signature of the collective response in density (Figure \ref{fig:contrast}).
    
\end{itemize}
The methodology developed in this work is a prelude to what will be possible with forthcoming photometric surveys such as the Chinese Space Station Telescope wide imaging survey \citep[CSST,][]{csst, cao+2018}, the Large Synoptic Survey Telescope (LSST, \citealt{lsst}), and Euclid space-based survey \citep{euclid}. CSST will cover 17,500 ${\rm deg}^2$ ($\approx 40\%$ of the sky), observe in the $NUV$,$ugriz$ photometric bands, reaching a limiting magnitude of $r\approx 26.0$ and $r\approx 27.2$ for the wide and deep imaging survey \citep[][]{qu+2023}, respectively. {The} LSST main survey will cover 30,000 ${\rm deg}^2$ and reach up to $r \approx 27.5$ at the end of its mission. The depth to be reached by these surveys, with lower magnitude errors compared to LSDR9, will allow us to probe the stellar halo fraction at a higher spatial angular resolution and larger distances. This will refine the measurement of the gravitational wake width in the MW halo due to the LMC, as shown in Figure \ref{fig:dens-lmc-orbit}, particularly when extending this work's methodology to Main Sequence Turn Off stars. These stars are not only more numerous than BHB stars \citep[e.g.][]{bell+2010}, but they will reduce the Poisson noise and allow for finer measurements of density variations across/along the wake and across overdensities \citep{sanderson+2019}. \par
Finally, we present in Appendix \ref{app:catalog} a catalogue of BHB star candidates with their respective probabilities calculated from Equation \ref{eq:Pgrz}.

\section*{Acknowledgements}

{The authors wish to thank the anonymous referee for their comments which helped improve the quality of this work.} J.A. \& C.L. acknowledge funding from the European Research Council (ERC) under the European Union’s Horizon 2020 research and innovation programme (grant agreement No. 852839). SK acknowledges support from the Science \& Technology Facilities Council (STFC) grant ST/Y001001/1.  J.A. thanks Guilherme Limberg, Hélio Perottoni, Friedrich Anders, and Matt Orkney for conversations that contributed to improve the quality of this work.

This paper made use of the Whole Sky Database (wsdb) created and maintained by Sergey Koposov at the Institute of Astronomy, Cambridge with financial support from the Science \& Technology Facilities Council (STFC) and the European Research Council (ERC).

Some of the results in this paper have been derived using the {\tt healpy} and {\tt HEALPix} packages.

The Legacy Surveys consist of three individual and complementary projects: the Dark Energy Camera Legacy Survey (DECaLS; Proposal ID \#2014B-0404; PIs: David Schlegel and Arjun Dey), the Beijing-Arizona Sky Survey (BASS; NOAO Prop. ID \#2015A-0801; PIs: Zhou Xu and Xiaohui Fan), and the Mayall z-band Legacy Survey (MzLS; Prop. ID \#2016A-0453; PI: Arjun Dey). DECaLS, BASS and MzLS together include data obtained, respectively, at the Blanco telescope, Cerro Tololo Inter-American Observatory, NSF’s NOIRLab; the Bok telescope, Steward Observatory, University of Arizona; and the Mayall telescope, Kitt Peak National Observatory, NOIRLab. Pipeline processing and analyses of the data were supported by NOIRLab and the Lawrence Berkeley National Laboratory (LBNL). The Legacy Surveys project is honored to be permitted to conduct astronomical research on Iolkam Du’ag (Kitt Peak), a mountain with particular significance to the Tohono O’odham Nation.

NOIRLab is operated by the Association of Universities for Research in Astronomy (AURA) under a cooperative agreement with the National Science Foundation. LBNL is managed by the Regents of the University of California under contract to the U.S. Department of Energy.

This project used data obtained with the Dark Energy Camera (DECam), which was constructed by the Dark Energy Survey (DES) collaboration. Funding for the DES Projects has been provided by the U.S. Department of Energy, the U.S. National Science Foundation, the Ministry of Science and Education of Spain, the Science and Technology Facilities Council of the United Kingdom, the Higher Education Funding Council for England, the National Center for Supercomputing Applications at the University of Illinois at Urbana-Champaign, the Kavli Institute of Cosmological Physics at the University of Chicago, Center for Cosmology and Astro-Particle Physics at the Ohio State University, the Mitchell Institute for Fundamental Physics and Astronomy at Texas A\&M University, Financiadora de Estudos e Projetos, Fundacao Carlos Chagas Filho de Amparo, Financiadora de Estudos e Projetos, Fundacao Carlos Chagas Filho de Amparo a Pesquisa do Estado do Rio de Janeiro, Conselho Nacional de Desenvolvimento Cientifico e Tecnologico and the Ministerio da Ciencia, Tecnologia e Inovacao, the Deutsche Forschungsgemeinschaft and the Collaborating Institutions in the Dark Energy Survey. The Collaborating Institutions are Argonne National Laboratory, the University of California at Santa Cruz, the University of Cambridge, Centro de Investigaciones Energeticas, Medioambientales y Tecnologicas-Madrid, the University of Chicago, University College London, the DES-Brazil Consortium, the University of Edinburgh, the Eidgenossische Technische Hochschule (ETH) Zurich, Fermi National Accelerator Laboratory, the University of Illinois at Urbana-Champaign, the Institut de Ciencies de l’Espai (IEEC/CSIC), the Institut de Fisica d’Altes Energies, Lawrence Berkeley National Laboratory, the Ludwig Maximilians Universitat Munchen and the associated Excellence Cluster Universe, the University of Michigan, NSF’s NOIRLab, the University of Nottingham, the Ohio State University, the University of Pennsylvania, the University of Portsmouth, SLAC National Accelerator Laboratory, Stanford University, the University of Sussex, and Texas A\&M University.

BASS is a key project of the Telescope Access Program (TAP), which has been funded by the National Astronomical Observatories of China, the Chinese Academy of Sciences (the Strategic Priority Research Program “The Emergence of Cosmological Structures” Grant \# XDB09000000), and the Special Fund for Astronomy from the Ministry of Finance. The BASS is also supported by the External Cooperation Program of Chinese Academy of Sciences (Grant \# 114A11KYSB20160057), and Chinese National Natural Science Foundation (Grant \# 12120101003, \# 11433005).

The Legacy Survey team makes use of data products from the Near-Earth Object Wide-field Infrared Survey Explorer (NEOWISE), which is a project of the Jet Propulsion Laboratory/California Institute of Technology. NEOWISE is funded by the National Aeronautics and Space Administration.

The Legacy Surveys imaging of the DESI footprint is supported by the Director, Office of Science, Office of High Energy Physics of the U.S. Department of Energy under Contract No. DE-AC02-05CH1123, by the National Energy Research Scientific Computing Center, a DOE Office of Science User Facility under the same contract; and by the U.S. National Science Foundation, Division of Astronomical Sciences under Contract No. AST-0950945 to NOAO.

For the purpose of open access, the author has applied a Creative
Commons Attribution (CC BY) licence to any Author Accepted Manuscript version
arising from this submission.\\

\bibliographystyle{aa} 
\bibliography{ref}

\appendix

\section{White dwarf contamination}\label{app:wd}
In this section, we investigate the possible white dwarf contamination in identifying BHB stars through our methodology described in Section \ref{sec:ccBHB}.

\begin{figure}
\centering

\includegraphics[width=\linewidth]{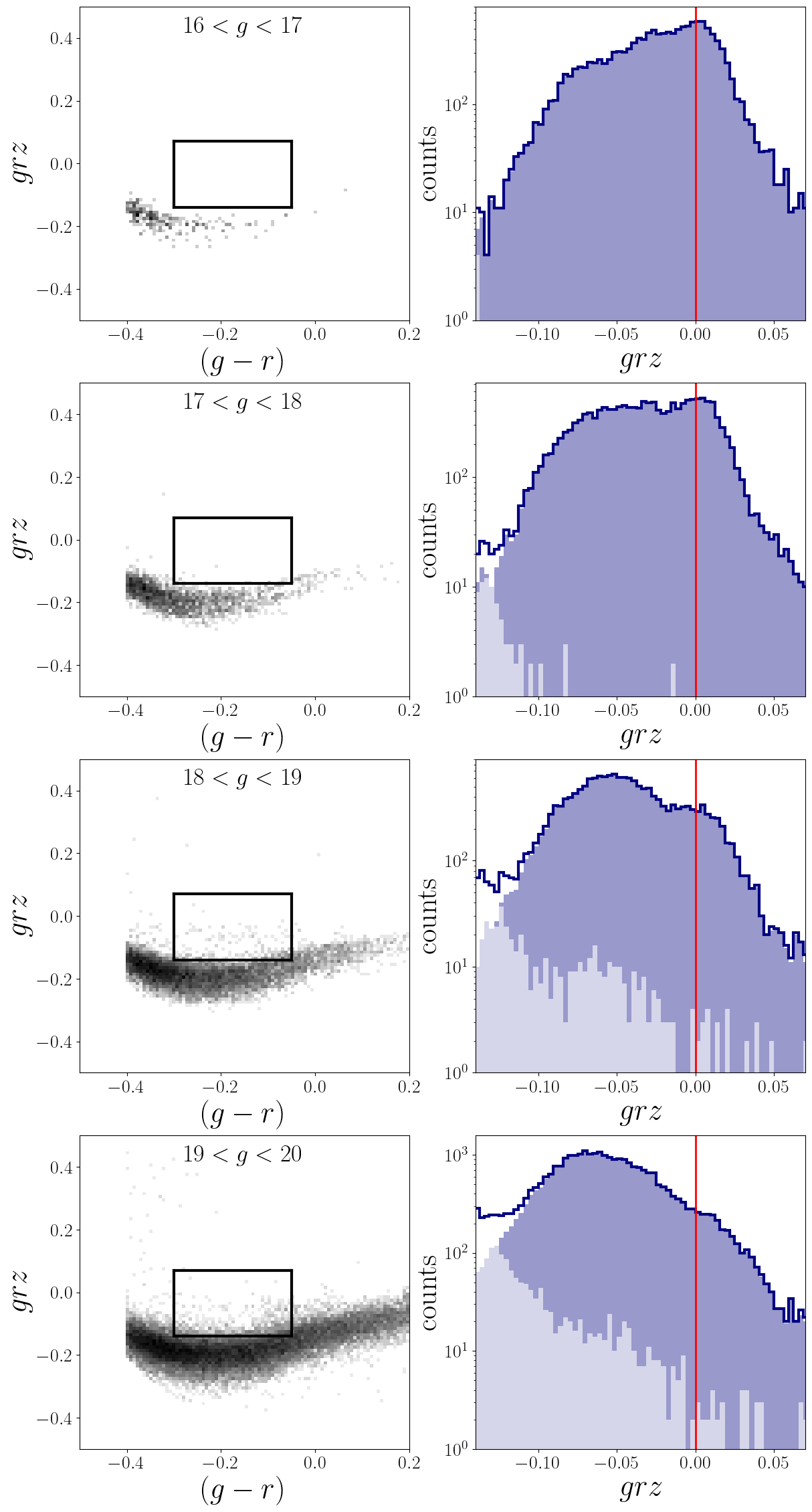}
    \caption{\textbf{Left column:} The $grz-(g-r)$ plane for the white dwarf stars in from the Gaia EDR3 catalogue \citep{fusillo+2021}. The black rectangle shows the region of interest in our study. \textbf{Right column:} The $grz$ histogram for objects inside the black rectangle. The white dwarf and non-white dwarf objects histograms are shown as the light and dark blue histograms, respectively. The solid blue line corresponds to their sum. The solid red line marks $grz=0$, the centre of the BHB distribution.}
    \label{fig:wd}
\end{figure}

We use the Gaia EDR3 white dwarf catalogue \citep{fusillo+2021} to verify whether this stellar population could contaminate the BHB fraction estimation. As we are not interested in the purity of a white dwarf sample, but rather its contamination, we used all their white dwarf candidates to crossmatch with our LSDR9 sample.  The left column in Figure \ref{fig:wd} shows the $grz-(g-r)$ plane, at different $g$ intervals, the white dwarf candidates. The black rectangle highlights the region of interest in our study and in the right column, we show the $grz$ histogram, in light blue, for the objects inside it. We also show, in blue, the histogram for all star objects excluding the white dwarfs, that lie inside the selected region. The solid blue line is the histogram of all objects in the selected $grz-(g-r)$ region. It is clear that for $g>17$ there is a low contamination fraction of white dwarf candidates in the expected BHB region.\par
\begin{figure}
\centering
\includegraphics{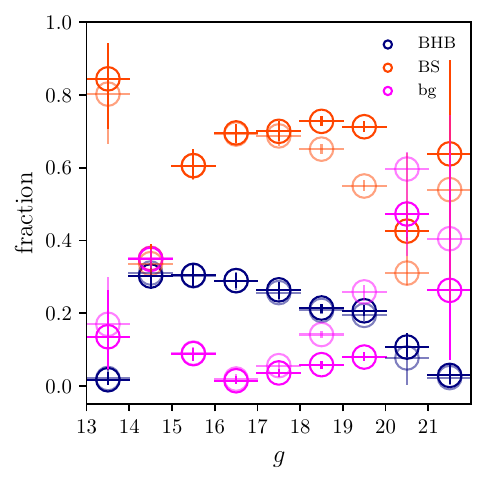}
    \caption{The fractions of each component from the MCMC modelling, described in Section \ref{sec:meth}, for objects with $-0.3<(g-r)<-0.05$ and $-0.14< grz < 0.07$ in the LSDR9-Gaia EDR3 crossmatch. The BHB Gaussian component has a hard prior for the mean and standard deviation. The dark/light markers are the fractions for a given $g$ interval when white dwarf candidates are excluded/included. The white dwarf contamination in the BHB fractions is negligible throughout $g$.}
    \label{fig:wdVSg}
\end{figure}

We verify the effect of not excluding white dwarf candidates on the BHB fraction estimation by fitting the $grz$ distribution for several intervals in $13<g<21$. We use the same methodology described in Section \ref{sec:meth} and also include strong prior in the BHB Gaussian parameters. Figure \ref{fig:wdVSg} shows the fractions for the BHB, BS and background components in blue, orange and magenta, respectively. The dark/light markers exclude/include white dwarf candidates. We can see that the BHB fraction is independent of the removal or not of white dwarf candidates, at least in the color-color regime of our study. The most noticeable effect is on the BS and background contamination. We, therefore, verify that white dwarfs are not significant sources of contamination in our study. 

\section{Radial density profiles}\label{app:hp-profile}

\begin{figure*}
\centering

\includegraphics[width=\linewidth]{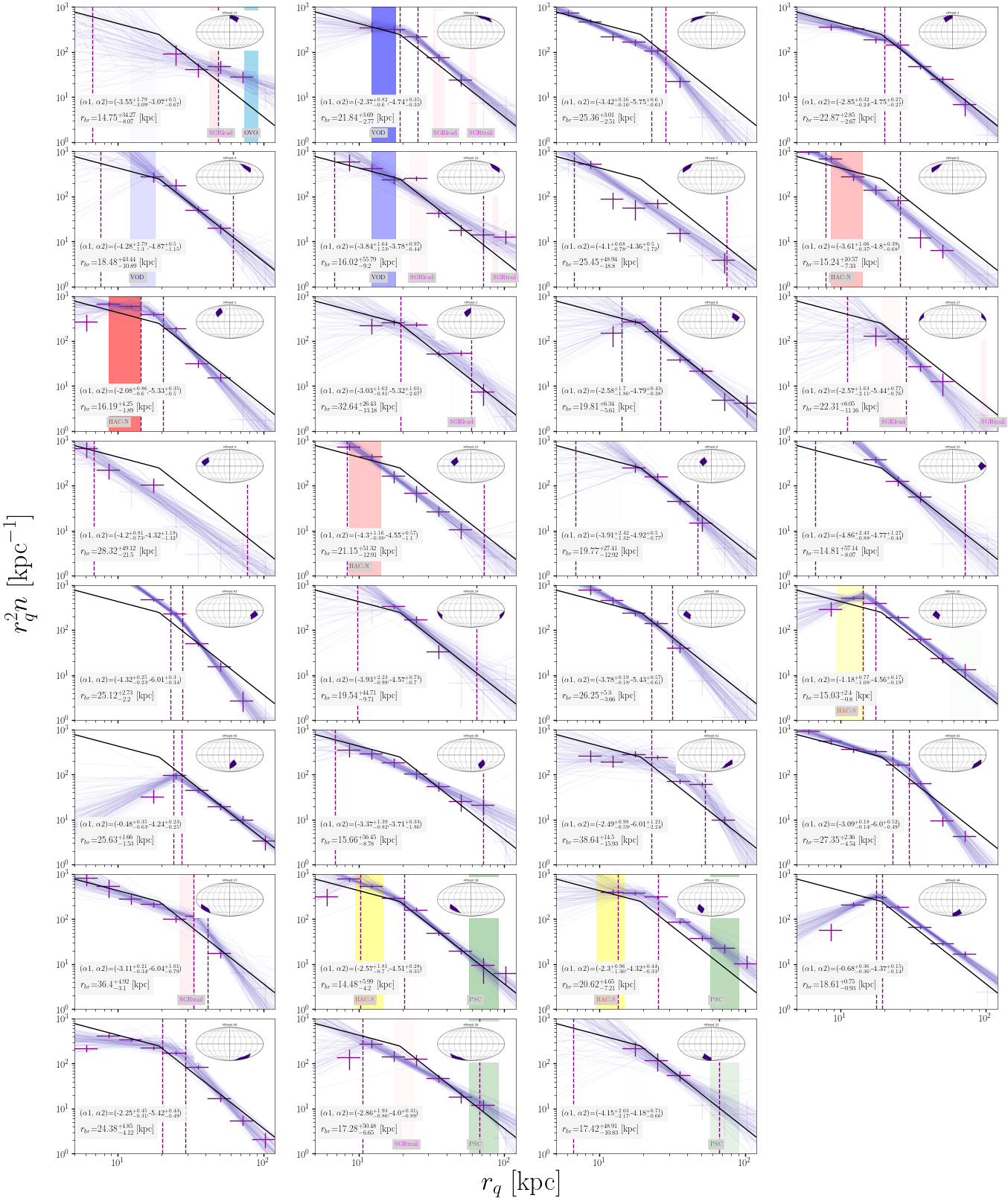}
    \caption{Radial density profiles for several patches in the Galactocentric sky as in Figure \ref{fig:rprofile-hp}. We show all the regions with more than four radial bins. See Section \ref{sec:prof-heal} for details.}
    \label{fig:panel}
\end{figure*}

In Section \ref{sec:prof-heal}, we divided the Galactocentric sky into 48 patches with the Python package {\tt healpy} \citep{healpix, healpy} using $nside=2$. Each region has an equal projected area of 859.44 ${\rm deg}^2$. We constructed the density profiles, see Section \ref{sec:methodology} for details, for 10 bins equally space in logarithm in the range $5 < r_{q}/{\rm kpc} < 120$. Figure \ref{fig:panel} shows the profiles of all the patches which have four or more bins in $r_{q}$.   \par
\begin{figure*}
\centering
\includegraphics[width=\linewidth]{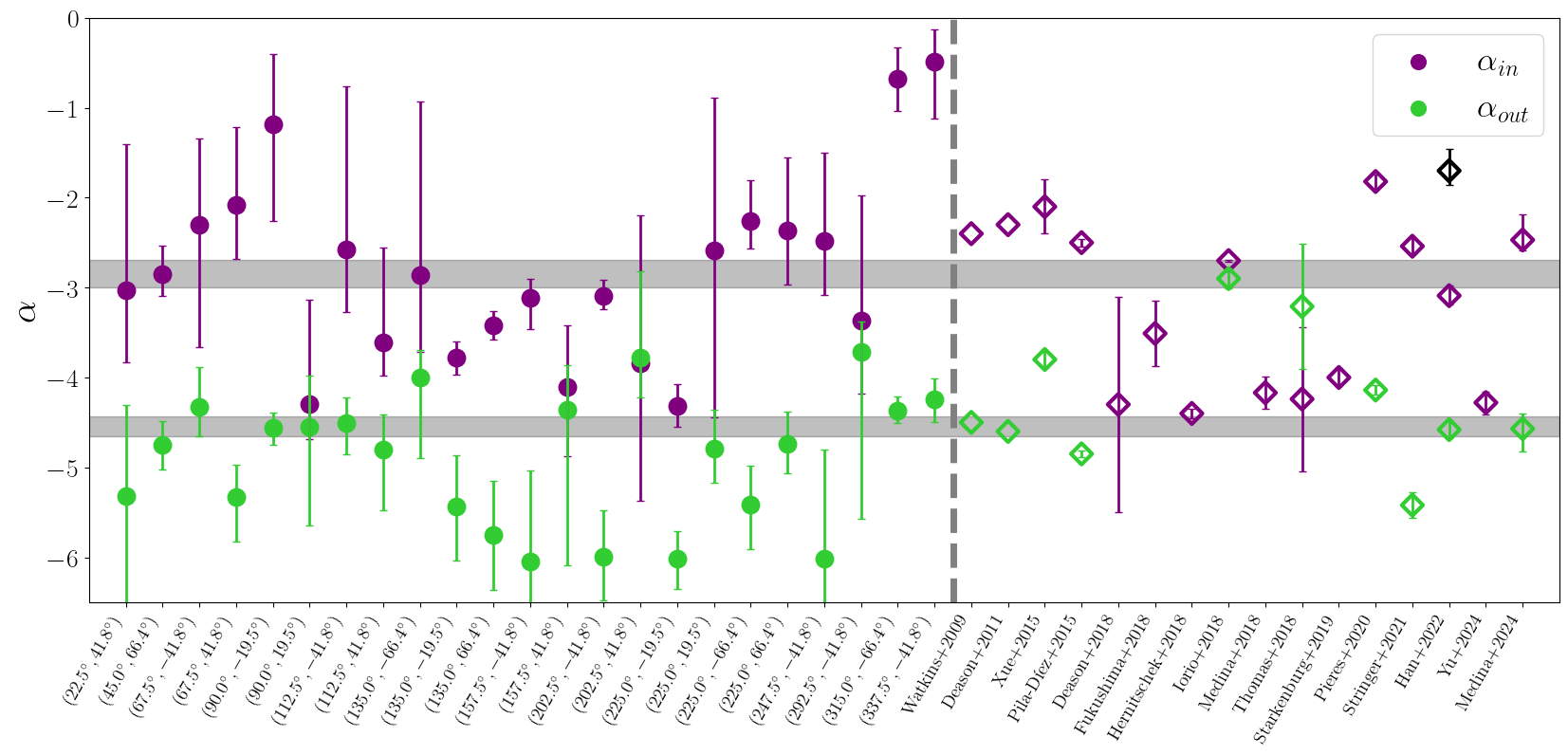}
    \caption{The inner (outer) slopes of the broken power law model fit of the density profiles shown in Figure \ref{fig:panel} shown as purple (green) circles. The x-axis shows the coordinates in degrees of the centre of the Galactocentric patch using {\tt healpy} Python package with HEALPIX pixelisation $nside=2$. The shaded region corresponds to the slopes measured for the average radial density profile as in Section \ref{sec:prof-ovd}. We also included the list compiled by \citet{yu+2024} of several works studying the halo radial density profile: \citet{watkins+2009, deason+2011, xue+2015, pila-diez+2015, deason+2018, fukushima+2018, hernitschek+2018, iorio+2017, medina+2018, thomas+2018, starkenburg+2019, pieres+2020,stringer+2021, han+2022, yu+2024, medina+2024}. These are shown as open diamonds. \citet{han+2022} favours a three-power-law model and thus the extra slope is shown as the black diamond. {The dashed vertical line separates data from our work from literature values.} } 
    \label{fig:alphas}
\end{figure*}

Figure \ref{fig:alphas} shows the inner and outer slopes, from the broken-power law fitting (see Section \ref{sec:rdp}) for each patch of the Galactocentric sky with four or more radial bins. The coordinates of the centre of each patch are given on the x-axis in degrees. The panel highlights the highly anisotropic stellar halo, as observed with BHB stars and discussed throughout in Section \ref{sec:discuss}. We also show the slopes obtained by several studies as compiled by \citet{yu+2024}. 

\section{LSDR9 BHB star candidates catalogue}\label{app:catalog}

We use objects from Legacy Survey Data Release 9 \citep{desi+2019} with the colour selection of Equation \ref{eq:cc}. We then construct a catalogue with 95446 objects containing LSDR9 {\tt release}, {\tt brickid} and {\tt objid}. These three parameters provide a unique identifier hash\footnote{\url{https://www.legacysurvey.org/dr9/catalogs/}}. The catalogue also includes R.A., Dec., the extinction corrected $g_0$, $r_0$, $z_0$, $grz$ (Equation \ref{eq:bhb}), {the absolute $M_g$ magnitude (Equation \ref{eq:Mgbhb}),} and the probability of a stellar object being a BHB star, $P_{BHB}$, calculated with Equation \ref{eq:Pgrz}. Table \ref{table:catalog} shows the header and first 10 rows of the catalogue available online. \footnote{\url{https://zenodo.org/records/12155488}}.

\begin{table*}

\begin{tabular}{ |p{1.cm} p{1cm} p{1cm} p{1cm} p{1cm} p{1.cm} p{1.cm}p{1.cm} p{1.cm} p{1.cm} p{1.cm}| }
 {\tt release} & {\tt brickid} & {\tt objid} &  R.A. [deg] & Dec [deg] & $g_0$ & $r_0$ & $z_0$ & $grz$ & $M_g$ & $P_{BHB}$  \\
 \hline
9010&279546&378&211.46&-8.89&16.41&16.61&16.84&0.04&0.54&0.55\\[0.11cm]
9010&279546&399&211.47&-8.88&19.04&19.22&19.49&-0.03&0.5&0.44\\[0.11cm]
9010&282395&4350&211.54&-8.42&14.55&14.79&15.07&0.02&0.63&0.85\\[0.11cm]
9010&282396&3890&211.75&-8.54&15.74&15.99&16.29&0.01&0.66&0.93\\[0.11cm]
9010&283822&5900&211.79&-8.32&20.56&20.69&20.93&-0.05&0.46&0.15\\[0.11cm]
9010&283819&817&210.84&-8.34&15.64&15.86&16.15&-0.0&0.58&0.91\\[0.11cm]
9010&283819&3721&210.98&-8.33&16.21&16.28&16.48&-0.05&0.43&0.01\\[0.11cm]
9010&285246&4170&210.96&-7.93&20.62&20.79&20.99&0.03&0.49&0.85\\[0.11cm]
9010&286673&4635&210.73&-7.82&19.72&19.83&20.0&0.02&0.45&0.9\\[0.11cm]
9010&286674&1976&210.85&-7.68&16.22&16.34&16.56&-0.01&0.46&0.77\\[0.11cm]
\end{tabular}
\caption{First 10 rows of the catalogue constructed with LSDR9 objects. See Appendix \ref{app:catalog} for details.}
\label{table:catalog}
\end{table*}

\end{document}